\documentclass[12pt,smallextended,natbib,runningheads,epsfig]{article}
\usepackage{latexsym, amsfonts}
\usepackage{amssymb, amsmath}
 \usepackage{amsfonts}
\usepackage{graphicx}
\usepackage{adjustbox}
\usepackage{setspace} 
\usepackage{rotating}
\usepackage{tikz}
\usepackage[autostyle]{csquotes}
\newcommand{\be}{\begin{equation}}
\newcommand{\ee}{\end{equation}}
\newcommand{\bea}{\begin{eqnarray}}
\newcommand{\eea}{\end{eqnarray}}
\newcommand{\bean}{\begin{eqnarray*}}
\newcommand{\eean}{\end{eqnarray*}}
\newcommand{\brray}{\begin{array}}
\newcommand{\erray}{\end{array}}
\newcommand{\ben}{\begin{equation}{nonumber}}
\newcommand{\een}{\end{equation}{nonumber}}

\newtheorem{dfn}{Definition}[section]
\newtheorem{thm}[dfn]{Theorem}
\newtheorem{lmma}[dfn]{Lemma}
\newtheorem{ppsn}[dfn]{Proposition}
\newtheorem{crlre}[dfn]{Corollary}
\newtheorem{xmpl}[dfn]{Example}
\newtheorem{rmrk}[dfn]{Remark}

\newcommand{\bdfn}{\begin{dfn}}
\newcommand{\bthm}{\begin{thm}}
\newcommand{\blmma}{\begin{lmma}}
\newcommand{\bppsn}{\begin{ppsn}}
\newcommand{\bcrlre}{\begin{crlre}}
\newcommand{\bxmpl}{\begin{xmpl}}
\newcommand{\brmrk}{\begin{rmrk}}
\newcommand{\edfn}{\end{dfn}}
\newcommand{\ethm}{\end{thm}}
\newcommand{\elmma}{\end{lmma}}
\newcommand{\eppsn}{\end{ppsn}}
\newcommand{\ecrlre}{\end{crlre}}
\newcommand{\exmpl}{\end{xmpl}}
\newcommand{\ermrk}{\end{rmrk}}

\def\a*{{\cal A}_{h,*}}
\def\B{{\cal B}(h)}
\def\B1{{\cal B}_1(h)}
\def\b{{\cal B}^{\rm s.a.}(h)}
\def\b1{{\cal B}^{\rm s.a.}_1(h)}

\begin{document}
\begin{center}
{\large {\bf A Test for Multivariate Location Parameter in Elliptical Model based on Forward Search Method}}\\
{\large  Chitradipa Chakraborty$^{1}$, Subhra Sankar Dhar$^{2}$}\\
{\large IIT Kanpur, India$^{1,2}$ }\\
{Email: chitrac@iitk.ac.in$^{1}$, subhra@iitk.ac.in$^{2}$}\\
\end{center}


\begin{abstract}
In this article, we develop a test for multivariate location parameter in elliptical model based on the forward search estimator for a specified scatter matrix.\ Here, we study the asymptotic power of the test under contiguous alternatives based on the asymptotic distribution of the test statistics under such alternatives.\ Moreover, the performances of the test have been carried out for different simulated data and real data, and compared the performances with more classical ones.
\end{abstract}
%




%

\section{Introduction}
The multivariate forward search method that is mainly concerned with detecting outliers and determining
their effect on models fitted to data. To be precise, the method is the idea of fitting a model containing
outliers to subsets of an increasing size. In this article, a test for multivariate location parameter in elliptical distribution for a specified scatter matrix is constructed, where the location parameter is estimated by
multivariate forward search method. In this context, we would like to mention that though the multivariate forward search estimator of location parameter is known concept in the literature (see Johansen and
Nielsen (2010)), we here briefly describe that estimator for a sake of completeness. In the construction of the estimator, it is defined at step $\gamma\in (0, 1)$ as 
$$
{\dot{\mbox{\boldmath $\mu$}}}_{\gamma,n} = \sum\limits_{i=1}^n\frac{\eta_{i,\gamma,n}}{S_{\gamma,n}} \textbf{y}_i,
$$
where $\eta_{i,\gamma,n}=I(Md^2_{i,n}\leq\delta^2_{\gamma,n})$ and $S_{\gamma,n}=\sum\limits_{i=1}^n\eta_{i,\gamma,n}$.\ Here $I(A) = 1$ if $A$ is true, and otherwise, equals zero.\ Among other notations, $
Md^2_{i,n}=(\textbf{y}_i-\mbox{\boldmath $\mu$}_{0})'\Sigma^{-1}(\textbf{y}_i-\mbox{\boldmath $\mu$}_{0}), \ i=1,...,n,
$ is the population Mahalanobis distances, where $\mbox{\boldmath $\mu$}_{0}$ is specified in null hypothesis (i.e., hence, it is known to us), and $\Sigma$ is the known scatter matrix. Further, $\delta^2_{\gamma,n}$ is the $\gamma$-th quantile among $Md^2_{i,n}, \ i=1,...,n$.\ It is to be noted that, we have $Md^2_{(1),n}<...<Md^2_{(n),n}$ with probability one since the observations are obtained from a continuous distribution.\ In view of this fact, one may in fact take $\delta^2_{\gamma,n} = Md^2_{(m),n}$, where $m=[n\gamma]$.\ We would here like to emphasize that \mbox{\boldmath $\mu$}$_{0}$ will be considered as the initial estimator for this forward search methodology at any step since \mbox{\boldmath $\mu$}$_{0}$ is specified to us. After having this estimator of the location parameter \mbox{\boldmath $\mu$}, one can now formulate the test statistic $T_{n}^{1}$ to test $H_{0}$ against $H_{1}$    as $T_{n}^{1} = \left|\left|\sqrt{n} ({\dot{\mbox{\boldmath $\mu$}}}_{\gamma,n} - \mbox{\boldmath $\mu$}_{0})\right|\right|^2$, which is nothing but the square of the Euclidean distance between ${\dot{\mbox{\boldmath $\mu$}}}_{\gamma,n}$ and $\mbox{\boldmath $\mu$}_{0}$.\ For a sake of completeness, we here define the elliptical distribution of a random variable $\textbf{Y}$, whose density function is of the form $$f_\textbf{Y}(\textbf{y})= k {\vert \Sigma \vert}^{-\frac{1}{2}} g((\textbf{y}-\mbox{\boldmath $\mu$})'\Sigma^{-1}(\textbf{y}-\mbox{\boldmath $\mu$})).$$ Here,  $\mbox{\boldmath $\mu$} \in\mathbb{R}^{d}$ is the unknown location parameter, and $\Sigma$ is the $d\times d$ known positive-definite scatter matrix.\ Among the others, $k$ is the normalizing constant, i.e., $k = \frac{\Gamma (\frac{d}{2})}{{\pi}^{\frac{d}{2}}} {\left[\int \limits_{0}^\infty x^{\frac{d}{2}-1} g(x)dx \right]}^{-1}$, and $g(.)$ is the density generator function such that $\int \limits_{0}^\infty x^{\frac{d}{2}-1} g(x)dx < \infty$ (see, e.g., Fang, Kotz and Ng (1989)).\ Our objective here is to propose a test for the location parameter $\mbox{\boldmath $\mu$}$ using the test statistic based on the multivariate forward search location estimator.\ In addition, as a toolkit, to measure the performance of the test, we study the performance of the test under contiguous (or local) alternatives.\ Non-technically speaking, the distribution associated with contiguous alternatives converge to the distribution under null hypothesis as the sample size tends to infinity.\ Section 3 describes this concept elaborately. 


Note that the asymptotic distribution of the test based on the sample mean is consistent only when the variance of the random variable associated with the marginal distribution is finite, and for the test based on co-ordinate wise median, one needs to assume a condition that the marginal density function is positive in the neighbourhood of the population median. However, unlike the test based on the mean and the co-ordinate wise median, one does not need to assume any moment based condition or any condition on the feature of the density function to apply the test based on the forward search estimator, and that is one of the significant advantage to constitute the test based on the forward search method. Also, note that the sample mean has asymptotic breakdown point $=0$, and the co-ordinate wise median achieves the highest asymptotic breakdown point $= 1/2$. In this context, we would like to emphasize that the choice of $\gamma$ $(= 1/2)$ also allows the highest possible value of asymptotic breakdown point $(=1/2)$ of the multivariate forward search estimator of location parameter (see Property 1 in Appendix B). This is also a reason to develop the test based on forward search estimator.

The rest of the article is organized as follows.\ In Section 2, we formulate the  test statistics based on the multivariate forward search estimator and other classical estimators of location parameter.\ In that section, we study the consistency properties of the tests along with  their finite sample performances.\ Section 3 investigates the asymptotic powers of the tests under contiguous alternatives.\ Some concluding remarks are discussed in Section 4.\ All technical details of the tests are provided in Appendix A, and Appendix B contains a few properties of multivariate forward search estimator of location parameter. 
\section{Formulation of the Test along with Properties and  Performances}

Let us first formulate the goodness of test formally.\ Suppose that ${\cal{Y}} = \{\textbf{y}_1,...,\textbf{y}_n\}$ of size n from an Elliptical distribution (see Introduction) having an unknown location parameter \mbox{\boldmath $\mu$} when the scatter matrix $\Sigma$ is known.\ In this article, we now want to test $H_{0}: \mbox{\boldmath $\mu$} = \mbox{\boldmath $\mu$}_0$ against the alternative $H_{1}: \mbox{\boldmath $\mu$} \neq \mbox{\boldmath $\mu$}_0$, where \mbox{\boldmath $\mu$}$_{0}$ is specified to us.\ 
In this article, as we indicated in the Introduction, one may formulate a test statistic (denote it as $T_{n}^{1}$) based on the multivariate forward search estimator of location parameter to test $H_{0}$ against $H_{1}$.

\vspace{0.1 in}

\noindent\textbf{Remark 2.1.} {\it It is here appropriate to mention that one can consider any other appropriate distance function in principle.\ The advantage of choosing the Euclidean distance is its easier mathematical tractability, and as a consequence of that, the asymptotic distribution of $T_{n}^{1}$ will have a nice form of weighted chi-squared distribution in view of the fact that the forward search location estimator converges to a multivariate normal distribution (see Lemma 1).\ This fact can be derived by an application of well-known orthogonal decomposition of multivariate normal distribution.\ Beside this mathematical tractability, this formulation ensures that the test statistic is invariant under rotational transformation.\ In other words, for a given data ${\cal{X}} = \{{\bf x}_{1}, \ldots, {\bf x}_{n}\}$, $T_{n}^{1}({\cal{X}}) = T_{n}^{1}({\cal{Y}})$, where ${\cal{Y}} = \{A{\bf x}_{1}, \ldots, A{\bf x}_{n}\}$, and $A$ is an orthogonal matrix.}

\noindent We now state a theorem describing the asymptotic behaviour of the test based on $T_{n}^{1}$.

\vspace{0.1 in}

\noindent\textbf{Theorem 2.1.}
 {\it Let $c_{\alpha}$ be the $(1 - \alpha)$-th $(0 < \alpha < 1)$ quantile of the distribution of $\sum\limits_{i = 1}^{d}\lambda_{i}Z_{i}^{2}$, where $\lambda_{i}$s are the eigen values of $\Sigma_{1} = \left[\frac{1}{d\gamma} \frac{\pi^\frac{d}{2}}{\Gamma{(\frac{d}{2}})}\int \limits_{0}^\infty x^{\frac{d}{2}} g(x)dx\right]\Sigma$, and $Z_{i}$'s are the i.i.d.\ $N(0, 1)$ random variables.\ A test, which rejects $H_{0}$ when $T_{n}^{1} > c_{\alpha}$, will have asymptotic size $\alpha$.\ Further, such a test will be a consistent test in the sense that, when $H_{1}$ is true, the asymptotic power of the test will be one.}

To implement this test, one needs to compute the eigen values of $\Sigma_{1}$, which is straightforward since $\Sigma_{1}$ is known to us, and then in principle, one has to compute the $(1 - \alpha)$-th quantile of the weighted chi-squared distribution, where weights are the eigen values of $\Sigma_1$.\ However, the exact computation of a specified quantile from the weighted chi-squared distribution may not be easily tractable.\ To overcome this problem, one may generate a large sample from the weighted chi-squared distribution and empirically estimate the specified quantile.\ To compute the power also, one may generate a large sample repeatedly from the weighted non-central chi-squared distribution, and the proportion of $T_{n}^{1} > \hat{c}_{\alpha}$ will be the estimated power, where $\hat{c}_{\alpha}$ is the estimated critical value.  
\subsection{{\bf Consistency Properties of Other Three Tests}}
\label{sec:2}
As we have already seen that the test based on forward search estimator is consistent, it is now of interest to study the consistency  properties of the test based on other three well-known estimators.\ Since the sample mean, the co-ordinate wise median and the Hodges-Lehmann estimator are the most well-known estimators of multivariate location parameter, we here formulate three test statistics based on those estimators.\ Let us denote {\mbox{\boldmath $\hat{\mu}$}$_{SM}$ , \mbox{\boldmath $\hat{\mu}$}$_{CM}$ and \mbox{\boldmath $\hat{\mu}$}$_{HL}$ be the sample mean, the co-ordinate wise median and the Hodges-Lehmann estimator, respectively.\ 
The test statistics for the sample mean, the co-ordinate wise median and Hodges-Lehmann estimator based tests are $T_{n}^{2} =  \left|\left|\sqrt{n} (\mbox{\boldmath $\hat{\mu}$}_{SM} - \mbox{\boldmath $\mu$}_{0})\right|\right|^2$, $T_{n}^{3} = \left|\left|\sqrt{n}(\mbox{\boldmath $\hat{\mu}$}_{CM} - \mbox{\boldmath $\mu$}_{0})\right|\right|^2$ and $T_{n}^{4} = \left|\left|\sqrt{n}(\mbox{\boldmath $\hat{\mu}$}_{HL} - \mbox{\boldmath $\mu$}_{0})\right|\right|^2$, respectively, where $\mbox{\boldmath $\mu$}_{0}$ is specified in the null hypothesis as we mentioned earlier.\ 
In the following propositions, the asymptotic behaviour of the tests based on $T_{n}^{2}$, $T_{n}^{3}$ and $T_n^4$ will be described.\ For notational convenience, we denote $\sigma_{2}^{2}= Var(Y_{1})$, where $Y_{1}$ is the first component of the random vector ${\bf Y}$, $\sigma_{3}^{2} =  \frac{1}{4{g_1}^2 (0)}$, and  $\sigma_{4}^{2} =  \frac{1}{12 [\int{g}_1^2 (x)dx]^2}$ (see Hodges and Lehmann (1963)), where  $g_{1}(.)$ is the marginal density function of the first component of the random vector, whose probability density function is $f(.)$.\ Note that $\sigma_{2}^{2} \Sigma$, $\sigma_{3}^{2} \Sigma$ (see Babu and Rao (1988)) and $\sigma_{4}^{2} \Sigma$  are asymptotic variances of the sample mean, the co-ordinate wise median and the Hodges-Lehmann estimator after appropriate normalization, respectively.

The next propositions describe the consistency of the tests based on the sample mean, the co-ordinate wise median and Hodges-Lehmann estimator. 

\vspace{0.1 in}

\noindent\textbf{Proposition 2.1.}
{\it  Let $c^{*}_{\alpha}$ be the $(1 - \alpha)$-th $(0 < \alpha < 1)$ quantile of the distribution of $\sum\limits_{i = 1}^{d}\lambda^{*}_{i}Z_{i}^{*2}$, where $Z^{*}_{i}$'s are the i.i.d.\ $N(0, 1)$ random variables and $\lambda^{*}_{i}$s are the eigen values of $\sigma_{2}^{2}\Sigma$.\ A test, which rejects $H_{0}$ when $T_{n}^{2} > c^{*}_{\alpha}$, will have asymptotic size $\alpha$. Further, such a test will be a consistent test in the sense that, when $H_{1}$ is true, the asymptotic power of the test will be one.} 

\vspace{0.1 in}

\noindent\textbf{Proposition 2.2.} {\it  Let $c^{**}_{\alpha}$ be the $(1 - \alpha)$-th $(0 < \alpha < 1)$ quantile of the distribution of $\sum\limits_{i = 1}^{d}\lambda^{**}_{i}Z_{i}^{**2}$, where $Z^{**}_{i}$'s are the i.i.d.\ $N(0, 1)$ random variables and $\lambda^{**}_{i}$s are the eigen values of $\sigma_{3}^{2}\Sigma$.\ A test, which rejects $H_{0}$ when $T_{n}^{3} > c^{**}_{\alpha}$, will have asymptotic size $\alpha$. Further, such a test will be a consistent test in the sense that, when $H_{1}$ is true, the asymptotic power of the test will be one. }

\vspace{0.1 in}

\noindent\textbf{Proposition 2.3.} {\it  Let $c^{***}_{\alpha}$ be the $(1 - \alpha)$-th $(0 < \alpha < 1)$ quantile of the distribution of $\sum\limits_{i = 1}^{d}\lambda^{***}_{i}Z_{i}^{***2}$, where $Z^{***}_{i}$'s are the i.i.d.\ $N(0, 1)$ random variables and $\lambda^{**}_{i}$s are the eigen values of $\sigma_{4}^{2}\Sigma$.\ A test, which rejects $H_{0}$ when $T_{n}^{4} > c^{***}_{\alpha}$, will have asymptotic size $\alpha$. Further, such a test will be a consistent test in the sense that, when $H_{1}$ is true, the asymptotic power of the test will be one. }

The assertion in Theorem 2.1. and Propositions 2.1., 2.2. and 2.3. indicate that the tests based on ${\dot{\mbox{\boldmath $\mu$}}}_{\gamma,n}$, {\mbox{\boldmath $\hat{\mu}$}$_{SM}$, \mbox{\boldmath $\hat{\mu}$}$_{CM}$ and \mbox{\boldmath $\hat{\mu}$}$_{HL}$ are consistent.\ In other words, the power of all of them will converge to one as the sample size converges to infinity.\ Hence, the performances of the tests are comparable when the sample size is infinite.
\subsection{{\bf Special Case: High Dimensional Space}}
\label{sec:2}
As we indicated at the end of the last section, Propositions 2.1., 2.2. and 2.3. along with Theorem 2.1. asserted that the tests based on $T_{n}^{1}$,  $T_{n}^{2}$, $T_{n}^{3}$ and $T_{n}^{4}$ are asymptotically equivalent (i.e., all of them are consistent) for finite and fixed dimension. This fact motivated us to investigate the efficiency study of the estimators ${\dot{\mbox{\boldmath $\mu$}}}_{\gamma,n}$, {\mbox{\boldmath $\hat{\mu}$}$_{SM}$, \mbox{\boldmath $\hat{\mu}$}$_{CM}$ and \mbox{\boldmath $\hat{\mu}$}$_{HL}$ when dimension is arbitrarily large, i.e., $d\rightarrow\infty$. Let $e_{1}(d) = \frac{Var(Y_{1})}{\left[\frac{1}{d\gamma} \frac{\pi^\frac{d}{2}}{\Gamma{(\frac{d}{2})}}\int \limits_{0}^\infty x^{\frac{d}{2}} g(x)dx\right]}$, $e_{2}(d) = \frac{d\gamma\Gamma{(\frac{d}{2})}}{4{g_1}^2 (0)\left[ \pi^\frac{d}{2}\int \limits_{0}^\infty x^{\frac{d}{2}} g(x)dx\right]}$ and $e_{3}(d) = \frac{d\gamma\Gamma{(\frac{d}{2})}}{12 (\int g_1^2(x) dx)^2 \left[ \pi^\frac{d}{2}\int \limits_{0}^\infty x^{\frac{d}{2}} g(x)dx\right]}$be the asymptotic efficiency of ${\dot{\mbox{\boldmath $\mu$}}}_{\gamma,n}$ relative to the sample mean, the co-ordinate wise median and the Hodges-Lehmann estimator, respectively. The next theorem describes the behaviour of $e_{1}(d)$, $e_{2}(d)$ and $e_3(d)$ as $d\rightarrow\infty$. 

\vspace{0.1 in}

\noindent\textbf{Theorem 2.2.}
{\it For $d$-dimensional Gaussian distribution, $e_{1}(d) = \frac{\gamma}{{(2\pi)}^\frac{d}{2}}$, $e_{2}(d) = \frac{\gamma
{\pi}^{1-\frac{d}{2}}}{{2}^(1+\frac{d}{2})}$ and $e_3(d) = \frac{\gamma \pi^{1-\frac{d}{2}}}{3 \times {(2)}^\frac{d}{2}}$, and consequently, in this case, $\displaystyle\lim_{d\rightarrow\infty} e_{1}(d) = 0$, $\displaystyle\lim_{d\rightarrow\infty} e_{2}(d) = 0$ and $\displaystyle\lim_{d\rightarrow\infty} e_{3}(d) = 0$. For $d$-dimensional Cauchy distribution, $e_{1} (d) = \infty$, $e_{2}(d) = \frac{\gamma d {\pi}^{\frac{3-d}{2}} \Gamma({\frac{d+1}{2}})}{4}$ and $e_{3}(d) = \frac{\gamma d {\pi}^{\frac{3-d}{2}} \Gamma({\frac{d+1}{2}})}{12}$, and consequently, in this case, $\displaystyle\lim_{d\rightarrow\infty} e_{1}(d) = \infty$, $\displaystyle\lim_{d\rightarrow\infty} e_{2}(d) = \infty$ and $\displaystyle\lim_{d\rightarrow\infty} e_{3}(d) = \infty$. For $d$-dimensional Spherical distribution with $g(x) = e^{-x^{100}}$, $e_{1} (d) = \frac{100 d\gamma \Gamma(\frac{d}{2})}{\pi^{\frac{d}{2}} \Gamma(\frac{1}{100}(\frac{d}{2}+1))}$, $e_{2}(d) = \frac{d\gamma \Gamma(\frac{d}{2}) (\Gamma(\frac{1}{200}))^2}{400 \pi^{\frac{d}{2}} \Gamma(\frac{1}{100}(\frac{d}{2}+1))}$ and $e_{3}(d) = \frac{53188.48 d\gamma \Gamma(\frac{d}{2})}{ \pi^{\frac{d}{2}} \Gamma(\frac{1}{100}(\frac{d}{2}+1))} $, and consequently, in this case, $\displaystyle\lim_{d\rightarrow\infty} e_{1}(d) = \infty$, $\displaystyle\lim_{d\rightarrow\infty} e_{2}(d) = \infty$ and $\displaystyle\lim_{d\rightarrow\infty} e_{3}(d) = \infty$.}

\vspace{0.1 in}

\noindent\textbf{Remark 2.2.}
{\it Theorem 2.2. indicates that why one can use forward search estimator for high dimensional data when the data is obtained from a heavy tailed distribution. We would here like to mention that the data generated from a heavy tailed distribution are more likely to have outliers, and hence, it is expected that forward search estimator is robust against the outliers even for high dimensional data obtained from a heavy tailed distribution. Moreover, for a \enquote{very light-tailed} distribution like Spherical  distribution with $g(x) = e^{-x^{100}}$, the forward search estimator performs best in terms of efficiency among all four estimators. Although in case of  Gaussian distribution, the sample mean posses better efficiency, and this fact indicates that one may use the sample mean to develop different methodologies in statistical inference even in high dimensional space if the data is obtained from a Gaussian distribution. }

Though the assertion in Theorem 2.2. indicates that forward search estimator posses better efficiency for high dimensional data when the data is obtained from a heavy tailed distribution or very light tailed distribution, it is our interest to see how the tests based on ${\dot{\mbox{\boldmath $\mu$}}}_{\gamma,n}$, {\mbox{\boldmath $\hat{\mu}$}$_{SM}$, \mbox{\boldmath $\hat{\mu}$}$_{CM}$ and \mbox{\boldmath $\hat{\mu}$}$_{HL}$ perform for finite sample and finite dimension, which is studied in next subsection.
\subsection{{\bf Finite Sample Level and Power Study}}
\label{sec:2}  
As we indicated at the end of the Section 2.1, Propositions 2.1., 2.2. and 2.3. along with Theorem 2.1. asserted that the tests based on $T_{n}^{1}$,  $T_{n}^{2}$ , $T_{n}^{3}$ and $T_{n}^{4}$ are asymptotically equivalent (i.e., all of them are consistent) for a fixed alternative hypothesis.\ At the next level, we therefore would like to see how the test based on $T_{n}^{1}$ performs compared to the tests based on $T_{n}^{2}$, $T_{n}^{3}$ and $T_{n}^{4}$ for the finite sample sizes.\ To study the finite sample performances, we here carry out some simulation studies to compare the finite sample powers of the tests.\ We consider tests with nominal level $5\%$ and in order to estimate the power of a test, we use Monte-Carlo replications.\ We use 1000 replications with each replication consisting of a sample of size $n = 100$ from alternative distribution and compute the powers of the tests based on $T_{n}^{1}$, $T_{n}^{2}$, $T_{n}^{3}$ and $T_{n}^{4}$ as the proportion of times the values of the corresponding test statistic is larger than the respective critical value.\ The critical values are computed by the method described in the last paragraph before Section 2.1. 

\vspace{1.4 in}

\noindent {\bf Table 1:} The finite sample power of different tests for different values $\beta$ at 5\% level of significance when sample size $ = 100$. Here $\gamma = 1/2$. 

\begin{center}
\begin{adjustbox}{max width=\textwidth} 
\small
\begin{tabular}{ |c|c|c|c|c|c|c|c|c|c|c|c| } 
\hline
Distribution & \multicolumn{11}{|c|}{$H_{0} = N_{4}({\bf 0}, I_{4})$ and $H_{1} =  (1 - \beta)  N_{4}({\bf 0}, I_{4}) + \beta N_{4}({\bf 5}, I_{4})$}\\
\hline
$\beta$ & 0 & 0.1 & 0.2 & 0.3 & 0.4 & 0.5 & 0.6 & 0.7 & 0.8 & 0.9 & 1\\
\hline
Test based on $T_{n}^{1}$   & 0.05 & 0.452 & 0.593 & 0.619 & 0.742 & 0.892 & 0.978 & 1 & 1 & 1 & 1\\ 
\hline
Test based on $T_{n}^{2}$  & 0.05 & 0.695 & 0.972 & 1 & 1 & 1 & 1 & 1 & 1  & 1&1\\
\hline
Test based on $T_{n}^{3}$   & 0.05 & 0.43 & 0.783 & 0.925 & 1 & 1 & 1 & 1 & 1 & 1 &1\\ 
\hline
Test based on $T_{n}^{4}$   & 0.05 & 0.65 & 0.969 & 0.989 & 1 & 1 & 1 & 1 & 1 & 1 &1\\ 
\hline
Distribution & \multicolumn{11}{|c|}{$H_{0} = C_{4}({\bf 0}, I_{4})$ and $H_{1} =  (1 - \beta)  C_{4}({\bf 0}, I_{4}) + \beta C_{4}({\bf 5}, I_{4})$}\\
\hline
Test based on $T_{n}^{1}$   & 0.05 & 0.642 & 0.849 & 0.885 & 0.925 & 0.993 & 1&  1& 1 & 1 & 1\\ 
\hline
Test based on $T_{n}^{2}$  & 0.05 & 0.009 & 0.201 & 0.127 & 0.328 & 0.074 & 0.415 & 0.091 & 0.168 & 0.613 & 0.006\\
\hline
Test based on $T_{n}^{3}$   & 0.05 & 0.638 & 0.92 &  1 & 1 & 1& 1 & 1& 1& 1& 1\\ 
\hline
Test based on $T_{n}^{4}$   & 0.05 & 0.782 & 0.98 & 1 & 1 & 1& 1 & 1& 1& 1& 1\\ 
\hline
Distribution & \multicolumn{11}{|c|}{$H_{0} = S_{4}({\bf 0}, I_{4})$ and $H_{1} =  (1 - \beta) S_{4}({\bf 0}, I_{4}) + \beta S_{4}({\bf 5}, I_{4})$}\\
\hline
Test based on $T_{n}^{1}$   & 0.05 & 0.741 & 0.811 & 0.929 & 1 & 1 & 1&  1& 1 &1 & 1\\ 
\hline
Test based on $T_{n}^{2}$  & 0.05 & 0.562 & 0.671 & 0.752 & 0.843 & 0.968 & 1&  1& 1 &1 & 1\\
\hline
Test based on $T_{n}^{3}$  & 0.05 & 0.623 & 0.777 & 0.861 & 0.921 & 1 & 1&  1& 1 &1 & 1\\
\hline
Test based on $T_{n}^{4}$   & 0.05 & 0.685 & 0.795 & 0.893 & 0.956 & 1 & 1 & 1 & 1 & 1 &1\\ 
\hline
\end{tabular}
\end{adjustbox}
\end{center}

Let us now consider three distributions, namely, $d$-dimensional Standard Gaussian distribution, $d$-dimensional Standard Cauchy distribution with probability density function $f_C({\bf x}) = (\Gamma((d + 1)/2)/{\pi}^{d/2}\Gamma(1/2))(1 + {\bf x}^{T}{\bf x})^{-(d +1)/2}$ and $d$-dimensional Spherical  distribution with $g(x) = e^{-x^{100}}$. 
It is an appropriate place to mention that $d=4$ is considered in our numerical study throughout this article unless mentioned otherwise.\ Under $H_{0}$, we first generate the data from 4-dimensional standard Gaussian, Cauchy distributions and Spherical  distribution with $g(x) = e^{-x^{100}}$, and to compute the power, we consider the distributions of the forms $(1 - \beta) F + \beta G$, where $\beta\in[0, 1]$, $ F $ and $ G $ are the same distribution upto the location parameter. For instance, $F ({\bf x}) = H ({\bf x})$ and $G({\bf x}) = H({\bf x} - {\mbox{\boldmath $\mu$}})$, where $H$ is any proper distribution function, and {\mbox{\boldmath $\mu$} is the location parameter. 
To summarize, $F$ is the distribution under $H_{0}$, and  $(1 - \beta) F + \beta G$ is the distribution under $H_{1}$.\ We here consider $G$ as 4-dimensional Gaussian, Cauchy distributions and Spherical  distribution with $g(x) = e^{-x^{100}}$,
${\mbox{\boldmath $\mu$}} = {\bf 5} = (5, 5, 5, 5)^{'}$ and $\gamma = 1/2$ for multivariate forward search estimator.\  The choice of $\gamma$ $(= 1/2)$ here is legitimate since that allows the highest asymptotic breakdown point of the multivariate forward search estimator of location parameter.\ {\it For the breakdown property of the multivariate forward search location estimator, we refer the readers to Property 1 in Appendix B.} All results are reported in Table 1, and in the table, $N_{4}$, $C_{4}$ and $S_4$ denote $4$-dimensional Gaussian, Cauchy distribution and Spherical  distribution with $g(x) = e^{-x^{100}}$ (i.e., \enquote{very light-tailed} distribution), respectively.    

When $\beta = 0$, the estimated power is close to the pre-specified size of the test $ = 0.05$ in all the cases since  $(1 - \beta) F + \beta G$ coincides with $F$, i.e., the distribution under the null hypothesis. The figures in Table 1 and Figure 1 indicate that the test based on $T_{n}^{1}$ performs well compared to the test based on $T_{n}^{2}$ when data are obtained from heavy-tailed distribution like the mixture of Cauchy distributions.\ This phenomena is expected since the forward search location estimator is more robust than the sample mean. In the case of Gaussian distribution, the test based on $T_{n}^{2}$ and $T_{n}^{4}$ performs better than the tests based on $T_{n}^{1}$ and $T_{n}^{3}$. For the Spherical  distribution with $g(x) = e^{-x^{100}}$ (i.e., light-tailed distribution), $T_n^1$ performs best compared to other three test statistics. {\it This fact can be further verified by the finite sample efficiency of the forward search estimator compared to the sample mean, the co-ordinate wise median and the Hodges-Lehmann estimator (see Property 2 in Appendix B).}

\vspace{1.4 in}

\noindent{\bf Figure 1:} Finite sample power of different tests for various values of $\beta$ at 5\% level of significance.

\begin{center}
\includegraphics[width=6in,height=4.7in]{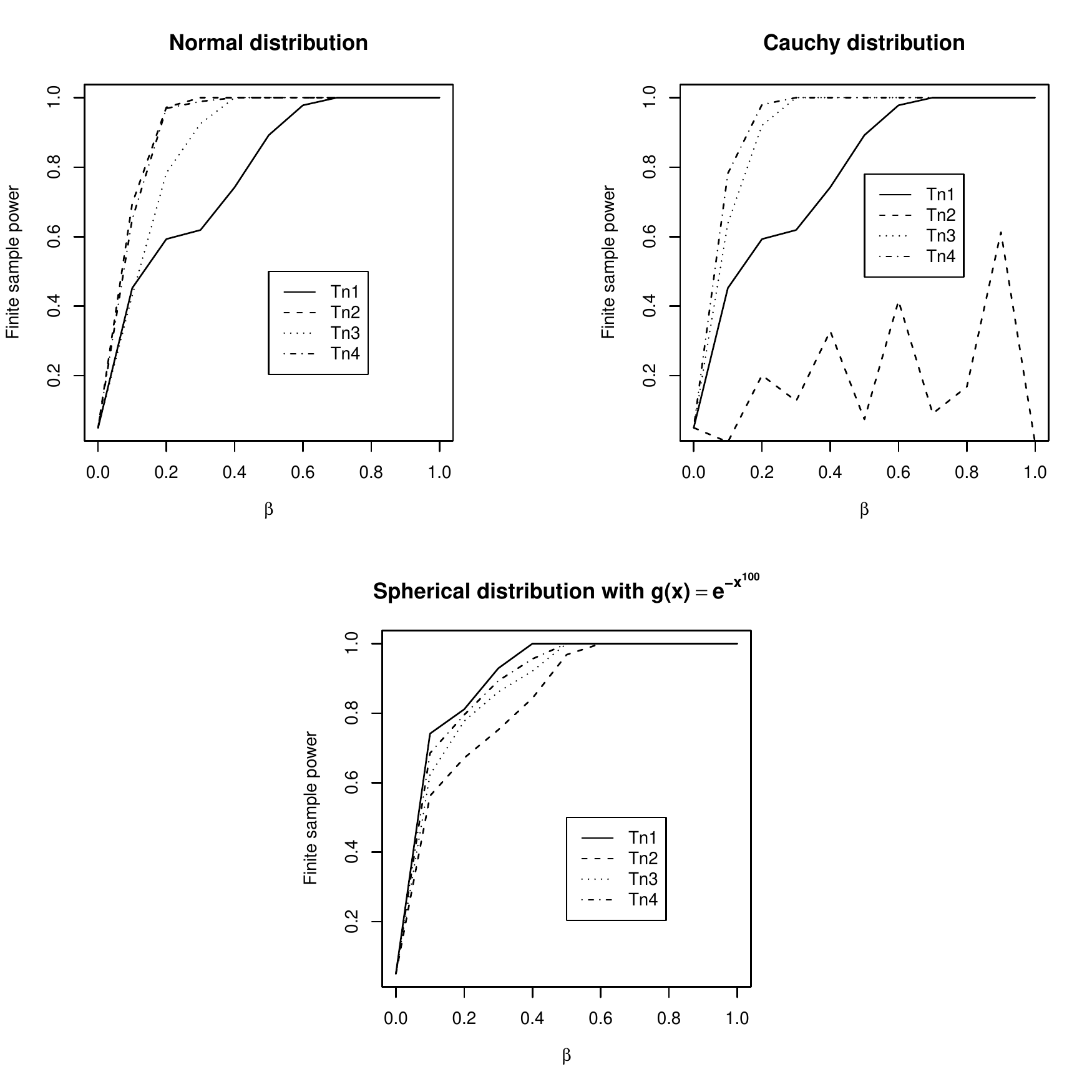}
\end{center}

\subsection{{\bf Real Data Analysis}}

{\bf Boston Housing Data:} This data set consists of 14 variables with size 506, and for details of the variables, we refer the readers at https://archive.ics.uci.edu/ml/datasets/Housing. In order to check the location parameter \mbox{\boldmath $\mu$} $= {\bf 0}$ (i.e., when $H_{0}$ is true) or not (i.e., when $H_{0}$ is not true), we carry out a bootstrap tests based on $T_{n}^{1}$, $T_{n}^{2}$, $T_{n}^{3}$ and $T_{n}^{4}$ and compute the $p$-values of the corresponding tests. We first compute the value of $T_{n}^{i}$, $i = 1, 2, 3, 4$ (denote it as $t_{0}^{i}$) from the given data and to estimate $P_{H_{0}}[T_{n}^{i} > t_{0}^{i}]$, (i.e., $p$-value) $i = 1, 2, 3, 4$, we generate $j$ many bootstrap resamples from the given data. Let $T_{n}^{i, k}$ denote the estimate of $T_{n}^{i}$ for $k$-th resample ($k = 1, \ldots, j$), and the $p$-value of the $i$-th test is defined as $\frac{\sum\limits_{k = 1}^{j}1_{\{T_{n}^{i, k} > t_{0}^{i}\}}}{j}$. In this numerical study, we choose $j = 10000$, and the $p$-values of the tests based on $T_{n}^{1}$, $T_{n}^{2}$, $T_{n}^{3}$ and $T_{n}^{4}$ are $0.5146$, $0.0928$, $0.5814$ and $0.6117$, respectively. It is indicated by these $p$-values, the test based on $T_{n}^{1}$, $T_{n}^{3}$ and $T_{n}^{4}$ do not reject the null hypothesis, i.e. in other words, the location parameter of the distribution associated with the data equals zero vector. On the other hand, the small p-value of the test based on $T_{n}^{2}$ rejects the null hypothesis. Here, we would like to mention that the result obtained from the mean based test is inconsistent with the original situation since this data contain large number of the outliers (see Figure 2). This real data analysis again establish the fact that the forward search estimator is robust against the outliers. 

\vspace{0.1 in}

\noindent{\bf Figure 2:} Plot of Boston data-set.

\begin{center}
\includegraphics[width=4.5in,height=3.8in]{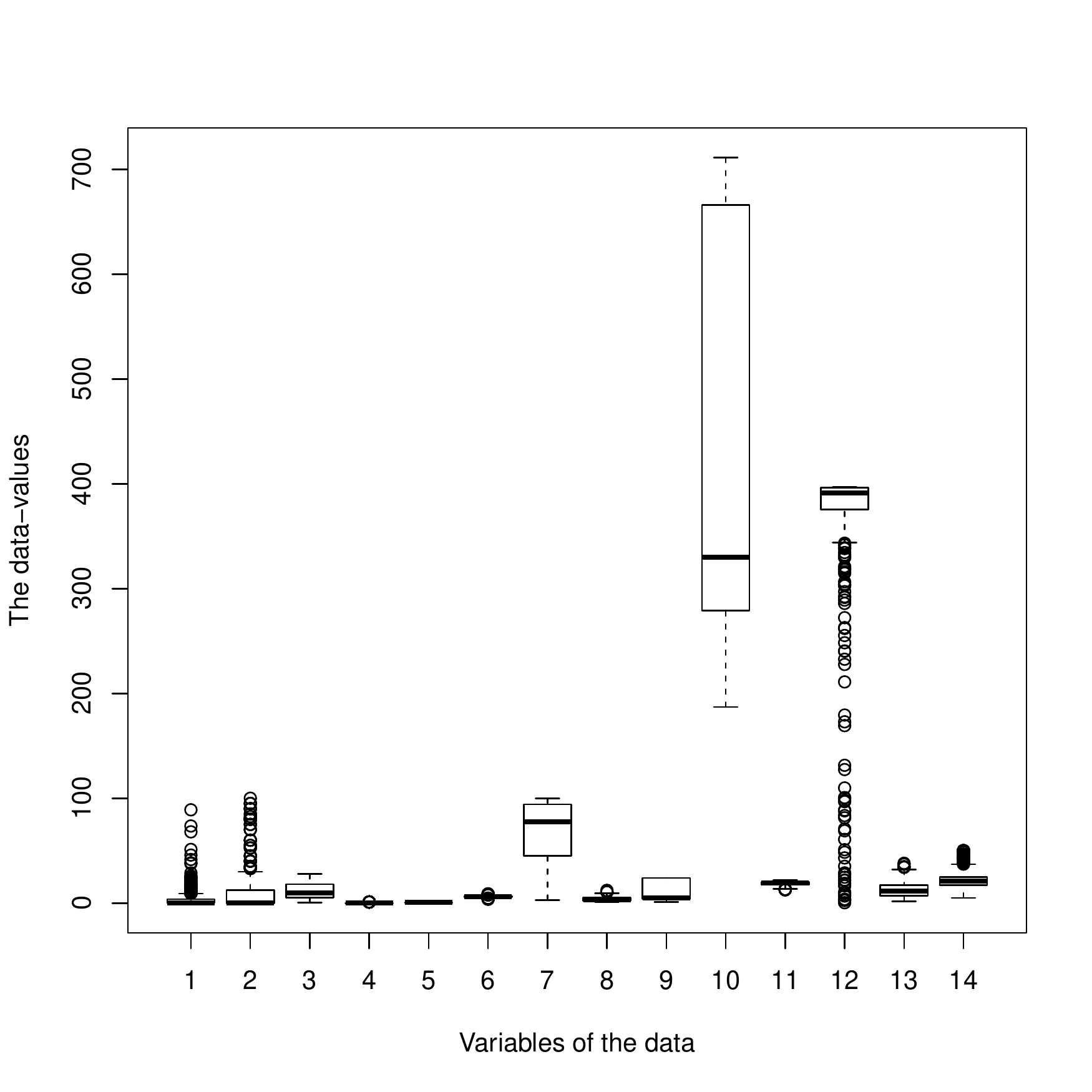}
\end{center}


\section{Asymptotic Power Study: Local Alternatives}
\label{sec:2}
As we have already seen that the tests based on $T_{n}^{1}$, $T_{n}^{2}$, $T_{n}^{3}$ and $T_{n}^{4}$ are consistent, we would now like to investigate the asymptotic power study of these tests under contiguous (or local) alternatives (see, e.g., Hajek, Sidak and Sen (1999)).\ Recently, the concept of contiguity has been described in Dhar, Dassios and Bergsma (2016); however, we again explain this concept here for sake of completeness.\  Precisely, the sequence of probability measures $Q_n$ is contiguous with respect to the sequence of probability measures $P_n$ if $P_n(A_n) \rightarrow 0$ implies that $Q_n(A_n) \rightarrow 0$ for every sequence of measurable sets $A_n$, where $(\Omega_n , A_n)$ is the sequence of measurable spaces, and $P_n$ and $Q_n$ are two probability measures defined on $(\Omega_n , A_n)$. Le Cam's first lemma (see, e.g., Hajek, Sidak and Sen (1999)) characterizes the contiguity based on the asymptotic distribution of the likelihood ratios between $P_{n}$ and $Q_{n}$. He established that the sequence $Q_n$ will be contiguous with respect to the sequence $P_n$ if $\log \frac{dQ_n}{dP_n}$ asymptotically follows the Gaussian distribution with mean = $-\frac{\sigma^2}{2}$ and variance = $\sigma^2$ under $P_n$ (see Hajek, Sidak and Sen (1999), p. 254, Corollary to Lecam's first lemma), where $\sigma > 0$ is a constant. Suppose that we now want to test $H_{0} : \mbox{\boldmath $\mu$} = \mbox{\boldmath $\mu$}_0$ against a sequence of alternatives $H_{1n} : \mbox{\boldmath $\mu$}_n = \mbox{\boldmath $\mu$}_0 + \frac{\mbox{\boldmath $\delta$}}{\sqrt{n}}, n=1,2,...$ for a fixed $\mbox{\boldmath $\delta$} = \{\delta_1, \ldots,\delta_d\}\in\mathbb{R}^{d}$. The contiguity of the distribution associated with $H_{1n}$ relative to the distribution associated with $H_{0}$ under certain condition has been stated in the following theorem.


\vspace{0.1 in}

\noindent\textbf{Theorem 3.1.}
{\it Let ${\bf Y}$ be a random vector associated with a distribution function $F(., \mbox{\boldmath $\mu$})$ having probability density function $f(., \mbox{\boldmath $\mu$})$. The probability density function $f(., \mbox{\boldmath $\mu$})$ is assumed to be twice continuously differentiable with respect to \mbox{\boldmath $\mu$}. Now, the sequence of distributions associated with $H_{1n}$ with respect to the distributions associated with $H_{0}$ will be contiguous when $E\left\{\frac{\partial^{2}}{\partial\mu_{i}\partial\mu_{j}}\log f({\bf y},  \mbox{\boldmath $\mu$}) \right\} < \infty$, where $\mu_{i}$ and $\mu_{j}$ are the $i$-th and the $j$-th components of \mbox{\boldmath $\mu$}, $i,j = 1\ldots n$.}

Note that the condition $E\left\{\frac{\partial^{2}}{\partial\mu_{i}\partial\mu_{j}}\log f({\bf y},  \mbox{\boldmath $\mu$}) \right\} < \infty$ will be satisfied for most of the well-known multivariate distribution functions. One can verify that this condition holds when data follow multivariate Gaussian and Cauchy distributions.\ In fact, the term $E\left\{\frac{\partial^{2}}{\partial\mu_{i}\partial\mu_{j}}\log f({\bf y},  \mbox{\boldmath $\mu$}) \right\}$ is essentially the $(i, j)$-th element of the information matrix with opposite sign. To summarize, Theorem 3.1. asserts that the sequence of distribution associated with $H_{1n}$ will be contiguous with respect to the distributions associated with $H_{0}$ when the determinant of the information matrix is finite.  We now state a theorem describing the asymptotic distributions of $T_{n}^{1}$, $T_{n}^{2}$, $T_{n}^{3}$ and $T_{n}^{4}$ under the contiguous alternatives $H_{1n}$. 

\vspace{0.1 in}

\noindent\textbf{Theorem 3.2.}
{\it Under $H_{1n}$ and the condition assumed in Theorem 3.1., $T_{n}^{1}$ converges weakly to $\sum\limits_{i = 1}^{d}\lambda_{i}(Z_{i} + a_{i})^{2}$, where $\lambda_{i}$s are the eigen values of 
$\Sigma_{1} =\left[\frac{1}{d\gamma} \frac{\pi^\frac{d}{2}}{\Gamma{(\frac{d}{2}})}\int \limits_{0}^\infty x^{\frac{d}{2}} g(x)dx\right]\Sigma$, $Z_{i}$'s are i.i.d.\ $N(0, 1)$ random variables, and $a_{i} = E\left\{({\dot{\mbox{\boldmath $\mu$}}}_{\gamma,n, i} - \mbox{\boldmath $\mu$}_{0, i})\sum\limits_{j = 1}^{d}\delta_{j}\frac{\partial g({\bf y}, \mbox{\boldmath $\mu$})}{\partial \mu_{j}}|_{\mbox{\boldmath $\mu$} = \mbox{\boldmath $\mu$}_{0}}\right\}$. Here for $i = 1, \ldots, d$, $\delta_{i}$, ${\dot{\mbox{\boldmath $\mu$}}}_{\gamma, n, i}$, \mbox{\boldmath $\mu$}$_{0, i}$ and $\mu_{i}$ are the $i$-th component of \mbox{\boldmath $\delta$}, ${\dot{\mbox{\boldmath $\mu$}}}_{\gamma, n}$, \mbox{\boldmath $\mu$}$_{0}$ and \mbox{\boldmath $\mu$}, respectively, and $g({\bf y}, \mbox{\boldmath $\mu$}) = \log f({\bf y}, \mbox{\boldmath $\mu$})$. Further, under those alternatives, $T_{n}^{2}$ converges weakly to $\sum\limits_{i = 1}^{d}\lambda^{*}_{i}(Z_{i}^{*} + a_{i}^{*})^{2}$, where $\lambda^{*}_{i}$s are the eigen values of $\sigma_{2}^{2}\Sigma,$  $Z^{*}_{i}$'s are i.i.d.\ $N(0, 1)$ random variables and $a_{i}^{*} = E\left\{(\mbox{\boldmath $\hat{\mu}$}_{SM, i} - \mbox{\boldmath $\mu$}_{0, i})\sum\limits_{j = 1}^{d}\delta_{j}\frac{\partial g({\bf y}, \mbox{\boldmath $\mu$})}{\partial \mu_{j}}|_{\mbox{\boldmath $\mu$} = \mbox{\boldmath $\mu$}_{0}}\right\}$. Here \mbox{\boldmath $\hat{\mu}$}$_{SM, i}$ is the $i$-th component of \mbox{\boldmath $\hat{\mu}$}$_{SM}$. Furthermore, under the same alternatives, $T_{n}^{3}$ converges weakly to $\sum\limits_{i = 1}^{d}\lambda^{**}_{i}(Z_{i}^{**} + a_{i}^{**})^{2}$, where $\lambda^{**}_{i}$s are the eigen values of $\sigma_{3}^{2}\Sigma$,  $Z^{**}_{i}$'s are i.i.d.\ $N(0, 1)$ random variables and $a_{i}^{**} = E\left\{(\mbox{\boldmath $\hat{\mu}$}_{CM, i} - \mbox{\boldmath $\mu$}_{0, i})\sum\limits_{j = 1}^{d}\delta_{j}\frac{\partial g({\bf y}, \mbox{\boldmath $\mu$})}{\partial \mu_{j}}|_{\mbox{\boldmath $\mu$} = \mbox{\boldmath $\mu$}_{0}}\right\}$. Here \mbox{\boldmath $\hat{\mu}$}$_{CM, i}$ is the $i$-th component of \mbox{\boldmath $\hat{\mu}$}$_{CM}$. Under the similar alternatives, $T_{n}^{4}$ also converges weakly to $\sum\limits_{i = 1}^{d}\lambda^{***}_{i}(Z_{i}^{***} + a_{i}^{***})^{2}$, where $\lambda^{***}_{i}$s are the eigen values of $\sigma_{4}^{2}\Sigma,$  $Z^{***}_{i}$'s are i.i.d.\ $N(0, 1)$ random variables and $a_{i}^{***} = E\left\{(\mbox{\boldmath $\hat{\mu}$}_{HL, i} - \mbox{\boldmath $\mu$}_{0, i})\sum\limits_{j = 1}^{d}\delta_{j}\frac{\partial g({\bf y}, \mbox{\boldmath $\mu$})}{\partial \mu_{j}}|_{\mbox{\boldmath $\mu$} = \mbox{\boldmath $\mu$}_{0}}\right\}$. Here \mbox{\boldmath $\hat{\mu}$}$_{HL, i}$ is the $i$-th component of \mbox{\boldmath $\hat{\mu}$}$_{HL}$, and $\sigma_{2}^{2}$, $\sigma_{3}^{2}$ and $\sigma_{4}^{2}$ are same as defined as before Propositions 2.1.}

\vspace{0.1 in}

\noindent\textbf{Remark 3.1.} {\it As discussed in the Introduction, the asymptotic distribution of $T_{n}^{2}$ is valid only when the variance of the random variable associated with the marginal distribution is finite, and for the test based on $T_{n}^{3}$, one needs to assume a condition that the marginal density function is positive in the neighbourhood of the population median. For the reason mentioned at the beginning of this remark, we will have asymptotic power $= 0$ for the test based on $T_{n}^{2}$ when the data is obtained from multivariate Cauchy distribution (see Table 2). 
 } 

The asymptotic power of the tests based on $T_{n}^{1}$, $T_{n}^{2}$, $T_{n}^{3}$ and $T_{n}^{4}$ under the contiguous alternatives $H_{1n}$ follow from the assertion in Theorem 3.2. Corollary 3.1. describes it.

\vspace{0.1 in}

\noindent\textbf{Corollary 3.1.} {\it Under $H_{1n}$ and the condition assumed in Theorem 3.2., the asymptotic power of the test based on $T_{n}^{1}$ is given by $P_{\mbox{\boldmath $\delta$}}\left[\sum\limits_{i = 1}^{d}\lambda_{i}(Z_{i} + a_{i})^{2} > c_{\alpha}\right]$, where $c_{\alpha}$ is such that $\displaystyle P_{\mbox{\boldmath $\delta$ = {\bf 0}}}\left[\sum\limits_{i = 1}^{d}\lambda_{i}(Z_{i} + a_{i})^{2} > c_{\alpha}\right] = \alpha$.\ Further, under those alternatives, the asymptotic power of the tests based on $T_{n}^{2}$, $T_{n}^{3}$ and $T_{n}^{4}$ are given by $P_{\mbox{\boldmath $\delta$}}\left[\sum\limits_{i = 1}^{d}\lambda_{i}^{*}(Z_{i}^{*} + a_{i}^{*})^{2} > c^{*}_{\alpha}\right]$, \\$P_{\mbox{\boldmath $\delta$}}\left[\sum\limits_{i = 1}^{d}\lambda_{i}^{**}(Z_{i}^{**} + a_{i}^{**})^{2} > c^{**}_{\alpha}\right]$ and $P_{\mbox{\boldmath $\delta$}}\left[\sum\limits_{i = 1}^{d}\lambda_{i}^{***}(Z_{i}^{***} + a_{i}^{***})^{2} > c^{***}_{\alpha}\right]$, where $c^{*}(\alpha)$, $c^{**}(\alpha)$ and $c^{***}(\alpha)$ are such that $P_{\mbox{\boldmath $\delta$} = {\bf 0}}\left[\sum\limits_{i = 1}^{d}\lambda_{i}^{*}(Z_{i}^{*} + a_{i}^{*})^{2} > c^{*}_{\alpha}\right] = \alpha$, $P_{\mbox{\boldmath $\delta$} = {\bf 0}}\left[\sum\limits_{i = 1}^{d}\lambda_{i}^{**}(Z_{i}^{**} + a_{i}^{**})^{2} > c^{**}_{\alpha}\right] = \alpha$ and $P_{\mbox{\boldmath $\delta$} = {\bf 0}}\left[\sum\limits_{i = 1}^{d}\lambda_{i}^{***}(Z_{i}^{***} + a_{i}^{***})^{2} > c^{***}_{\alpha}\right] = \alpha$.}

In order to compute the asymptotic power under the contiguous alternatives $H_{1n}$, one first needs to compute the critical value.\ It follows from the assertion in Corollary 3.1. that the critical value is essentially the $(1 - \alpha)$-th quantile of a certain weighted central chi-squared distribution.\ The exact computation of a certain quantile of the weighted chi-squared distribution is cumbersome.\ To avoid the direct computation, one may generate a large sample from the weighted central chi-squared distribution
and empirically estimate the specified quantile.\ Next, we try to compute $a_{i}$, $a_{i}^{*}$, $a_{i}^{**}$ and $a_{i}^{***}$ since these terms are involved in the expressions of the asymptotic power.\ In this step, for $i = 1, \ldots, d$, we estimate the expectations involved in $a_{i}$, $a_{i}^{*}$, $a_{i}^{**}$ and $a_{i}^{***}$ by the corresponding sample averages.\ Finally, to estimate the asymptotic power of the tests, we generate a large sample from the specified weighted non-central chi-squared distribution associated with the respective test and compute the proportion of the observations exceeding the corresponding estimated critical value.\ That proportion gives us the estimated asymptotic power. 

In Table 2, we provide the asymptotic power of the tests based on $T_{n}^{1}$, $T_{n}^{2}$, $T_{n}^{3}$ and $T_{n}^{4}$ for different values of $||\mbox{\boldmath $\delta$}||$, where $||.||$ is the Euclidean norm. In this study, the data are obtained from $4$-dimensional standard Gaussian, Cauchy distributions and Spherical  distribution with $g(x) = e^{-x^{100}}$ (for the expression of the probability density functions, see in Section 2.2.).\ To make the presentation concise, we assume that the components of \mbox{\boldmath $\delta$}, i.e., $\delta_{1}$, $\delta_{2}$, $\delta_{3}$ and $\delta_{4}$ are equal.

It is evident by the figures in Table 2 that the test based on the forward search estimator (i.e., the test based on $T_{n}^{1}$) performs better than the test based on the mean (i.e., the test based on $T_{n}^{2}$) when data are generated from a heavy tailed distribution.\ For Cauchy distribution, the performances of the tests based on $T_{n}^{1}$, $T_{n}^{3}$ and $T_n^4$ are comparable but the test based on $T_{n}^{2}$ fails to perform as we said above.\ In fact, as we stated in Remark 3.1., it is appropriate to emphasize that for the test based on $T_{n}^{2}$, the asymptotic power $= 0$ in this case.\ As it was expected, in the case of Gaussian distribution, the test based on $T_{n}^{2}$ (i.e., the test based on the sample mean) performs better than the tests based on $T_{n}^{1}$ (i.e., the test based on the forward search estimator) and $T_{n}^{3}$ (i.e., the test based on the co-ordinate wise median).\ The test based on $T_{n}^{4}$ is performing similar to the test based on $T_{n}^{2}$ for the normal distribution. For Spherical  distribution with $g(x) = e^{-x^{100}}$, $T_{n}^{1}$ performs better than all other three tests. {\it The aforementioned fact further can be verified by the asymptotic efficiency study of the forward search location estimator related to the sample mean, the co-ordinate wise median and the Hodges-Lehmann estimator (see Property 3 in Appendix B).}

\vspace{1.5in}

\noindent {\bf Table 2:} The asymptotic power of the tests based on $T_{n}^{1}$, $T_{n}^{2}$, $T_{n}^{3}$ and $T_{n}^{4}$ under $H_{1n}$ when data generated from $4$-dimensional standard Gaussian (denote it as $N_{4}(., .)$), Cauchy (denote it as $C_{4}(., .)$) and Spherical  with $g(x)=e^{-x^{100}}$ (denote it as $S_{4}(., .)$) distributions for different choices of $||\mbox{\boldmath $\delta$}||$ (i.e., $\delta_{1}$, $\delta_{2}$, $\delta_{3}$ and $\delta_{4}$). Here the level of significance $= 5\%$.

\begin{center}
\begin{adjustbox}{max width=\textwidth} 
\begin{tabular}{ |c|c|c|c|c|} 
\hline
Distribution & \multicolumn{4}{|c|}{$F = N_{4}(., I_{4})$}\\
\hline
Tests & Test based on $T_{n}^{1}$ & Test based on $T_{n}^{2}$ & Test based on $T_{n}^{3}$ & Test based on $T_{n}^{4}$\\
\hline
$||\mbox{\boldmath $\delta$}|| = 1$, $\delta_{1} = \delta_{2} = \delta_{3} = \delta_{4} = 0.5$ & 0.41 & 0.42 & 0.42 & 0.42\\
\hline 
$||\mbox{\boldmath $\delta$}|| = 1$, $\delta_{1} = \delta_{2} = \delta_{3} = \delta_{4} = -0.5$ & 0.52 & 0.53 & 0.53 & 0.53\\
\hline 
$||\mbox{\boldmath $\delta$}|| = 10$, $\delta_{1} = \delta_{2} = \delta_{3} = \delta_{4} = 5$ & 0.92 & 0.92& 0.91 & 0.92\\
\hline 
$||\mbox{\boldmath $\delta$}|| = 10$, $\delta_{1} = \delta_{2} = \delta_{3} = \delta_{4} = -5$ & 0.94 & 0.94 & 0.93 & 0.94\\
\hline
Distribution & \multicolumn{4}{|c|}{$F = C_{4}(., I_{4})$}\\
\hline
Tests & Test based on $T_{n}^{1}$ & Test based on $T_{n}^{2}$ & Test based on $T_{n}^{3}$ & Test based on $T_{n}^{4}$\\
\hline
$||\mbox{\boldmath $\delta$}|| = 1$, $\delta_{1} = \delta_{2} = \delta_{3} = \delta_{4} = 0.5$ & 0.96 & 0 & 0.95 & 0.96\\
\hline 
$||\mbox{\boldmath $\delta$}|| = 1$, $\delta_{1} = \delta_{2} = \delta_{3} = \delta_{4} = -0.5$ & 0.94 & 0 & 0.93 & 0.94\\
\hline 
$||\mbox{\boldmath $\delta$}|| = 10$, $\delta_{1} = \delta_{2} = \delta_{3} = \delta_{4} = 5$ & 0.98& 0 & 0.99 & 0.99\\
\hline 
$||\mbox{\boldmath $\delta$}|| = 10$, $\delta_{1} = \delta_{2} = \delta_{3} = \delta_{4} = -5$ & 0.97& 0& 0.98 & 0.98\\
\hline 
Distribution & \multicolumn{4}{|c|}{$F = S_{4}(., I_{4})$}\\
\hline
Tests & Test based on $T_{n}^{1}$ & Test based on $T_{n}^{2}$ & Test based on $T_{n}^{3}$ & Test based on $T_{n}^{4}$\\
\hline
$||\mbox{\boldmath $\delta$}|| = 1$, $\delta_{1} = \delta_{2} = \delta_{3} = \delta_{4} = 0.5$ & 0.93 & 0.89 & 0.9 & 0.91\\
\hline 
$||\mbox{\boldmath $\delta$}|| = 1$, $\delta_{1} = \delta_{2} = \delta_{3} = \delta_{4} = -0.5$ & 0.94 & 0.91 & 0.93 & 0.94\\
\hline 
$||\mbox{\boldmath $\delta$}|| = 10$, $\delta_{1} = \delta_{2} = \delta_{3} = \delta_{4} = 5$ & 0.98 & 0.91 & 0.94 & 0.96\\
\hline 
$||\mbox{\boldmath $\delta$}|| = 10$, $\delta_{1} = \delta_{2} = \delta_{3} = \delta_{4} = -5$ & 0.99 & 0.94 & 0.95 & 0.96\\
\hline 
\end{tabular}
\end{adjustbox}
\end{center}

\section{Concluding Remarks}
\label{sec:3}

{\bf Affine equivariant version of the test statistics:} Instead of $T_{n}^{1}$, $T_{n}^{2}$, $T_{n}^{3}$ and $T_{n}^{4}$,  one may consider affine equivariant versions $T_{n}^{1*} = n ||\Sigma^{-\frac{1}{2}}({\dot{\mbox{\boldmath $\mu$}}}_{\gamma, n} - \mbox{\boldmath $\mu$}_{0})||^2$, $T_{n}^{2*} = n ||\Sigma^{-\frac{1}{2}}(\mbox{\boldmath $\hat{\mu}$}_{SM} - \mbox{\boldmath $\mu$}_{0})||^2$, $T_{n}^{3*} = n ||\Sigma^{-\frac{1}{2}}(\mbox{\boldmath $\hat{\mu}$}_{CM} - \mbox{\boldmath $\mu$}_{0})||^2$ and $T_{n}^{4*} = n ||\Sigma^{-\frac{1}{2}}(\mbox{\boldmath $\hat{\mu}$}_{HL} - \mbox{\boldmath $\mu$}_{0})||^2$, respectively. However, since the scatter parameter $\Sigma$ is specified here, one can standardize the data based on the specified $\Sigma$, and this standardization procedure reduces $T_{n}^{i*}$ to $T_{n}^{i}$ for $i = 1, 2, 3$ and $4$. 

\noindent {\bf If $\Sigma$ is unknown:} One needs to estimate $\Sigma^{-1/2}$ to use $T_{n}^{i*}$, $i = 1, 2, 3, 4$ when $\Sigma$ is unknown. For $T_{n}^{1*}$, one may adopt the forward search methodology to estimate $\Sigma^{-\frac{1}{2}}$, and based on estimated $\Sigma^{-\frac{1}{2}}$, the behaviour of $T_{n}^{1*}$ remains an open problem. Note that the test based on $T_{n}^{2*}$ will coincide with the well-known Hotelling test (see, e.g., Hotelling (1931) and Puri and Sen (1971)) when $\Sigma^{-1/2}$ will be estimated by the variance-covariance matrix, and this test is {\it optimal} under Gaussian densities. However, as we have seen in Sections 2 and 3, the mean based test fails to perform well when data are obtained from heavy-tailed distribution since the mean is non-robust against the outliers. We would further like to point out that Hallin and Paindaveine (2002) proposed a test for multivariate location parameter based on the interdirection and the rank of psedu-Mahalanobish distance.\ In that article, they studied the asymptotic relative efficiency of their proposed tests under local alternatives relative to some other tests available in literature.\ In this article, we did not investigate the performance of the tests based on interdirection since the formulation of the tests are different from the tests based on $T_{n}^{1}$, $T_{n}^{2}$, $T_{n}^{3}$ and $T_{n}^4$. 

\noindent {\bf Non-elliptical distribution:} For non-elliptical distribution, the asymptotic distribution of the forward search estimator for the location parameter remains an open problem, whereas one can extend the results of the tests based on $T_{n}^{2}$ (i.e., the test based on the sample mean) and $T_{n}^{3}$ (i.e., the test based on the co-ordinate wise median) for non-elliptical distributions under some conditions. 

\noindent {\bf Main contribution of this article:} In this article, we propose a test for multivariate location parameter based on forward search method, which posses good power for heavy-tailed distribution.\ In addition, unlike the tests based on  $T_{n}^{2}$, $T_{n}^{3}$ and $T_n^4$, one neither needs to assume any moment condition nor needs to impose any restriction on the feature of the density function.\ Overall, our test has a rare phenomena that it can carried out under a mild condition but perform well even in the presence of outliers and/or large influential observations.

\section{Appendix A: Proofs}

In this section, we present the proofs of the theorems and some related lemmas. 

\vspace{0.1 in}

\noindent \textbf{Lemma 1:} As $n \rightarrow \infty$ under $H_0$, $\sqrt{n} ({\dot{\mbox{\boldmath $\mu$}}}_{\gamma,n} - \mbox{\boldmath $\mu$}_0)$ converges weakly to a $d$-dimensional normal distribution with the location parameter $= {\bf 0}$ and the scatter parameter $\Sigma_1$, where $\Sigma_{1} =\left[\frac{1}{d\gamma} \frac{\pi^\frac{d}{2}}{\Gamma{(\frac{d}{2})}}\int \limits_{0}^\infty x^{\frac{d}{2}} g(x)dx\right]\Sigma$.

\vspace{0.1 in}

\noindent \textbf{Proof of Lemma 1:} A straightforward application of polar transformation for an elliptical distribution and the construction of ${\dot{\mbox{\boldmath $\mu$}}}_{\gamma,n}$ along with the central limit theorem and Slutsky's theorem (see, e.g., Billingsley (1999)), it follows that $\sqrt{n}({\dot{\mbox{\boldmath $\mu$}}}_{\gamma,n}-\mbox{\boldmath $\mu$}_0)$ converges weakly to $d$-dimensional Gaussian distribution with mean $ = {\bf 0}$ and variance-covariance matrix $=  \left[\frac{1}{d\gamma} \frac{\pi^\frac{d}{2}}{\Gamma{(\frac{d}{2})}}\int \limits_{0}^\infty x^{\frac{d}{2}} g(x)dx\right]\Sigma$ under $H_0$.

\vspace{0.1 in}

\noindent \textbf{Proof of Theorem 2.1.:} To test $H_{0}: \mbox{\boldmath $\mu$} = \mbox{\boldmath $\mu$}_0$ against $H_{1}: \mbox{\boldmath $\mu$}\neq \mbox{\boldmath $\mu$}_0$, the power of the test based on $T_{n}^{1}$ is given by $P_{H_{1}}[T_{n}^{1} > c_{\alpha}]$, where $c_{\alpha}$ is the $(1 - \alpha)$-th $(0 < \alpha < 1)$ quantile of the distribution of $\sum\limits_{i = 1}^{d}\lambda_{i}Z_{i}^{2}$. Here, $\lambda_{i}$s are the eigen values of $\Sigma_{1} = \left[\frac{1}{d\gamma} \frac{\pi^\frac{d}{2}}{\Gamma{(\frac{d}{2}})}\int \limits_{0}^\infty x^{\frac{d}{2}} g(x)dx\right]\Sigma$, and $Z_{i}$s are the i.i.d.\ $N(0, 1)$ random variables. In view of the orthogonal decomposition of multivariate normal distribution, $T_{n}^{1}$ converges weakly to the distribution of  $\sum\limits_{i = 1}^{d}\lambda_{i}Z_{i}^{2}$, and hence, the asymptotic size of the test based on $T_{n}^{1}$ is $\alpha$. Let us now denote $\mbox{\boldmath $\mu$} = \mbox{\boldmath $\mu$}_{1} (\neq \mbox{\boldmath $\mu$}_0)$ under $H_{1}$, and we now consider


$\displaystyle\lim_{n\rightarrow\infty} P_{H_1} \left[T_{n}^{1} > c_{\alpha}\right]$ \\$= \displaystyle\lim_{n\rightarrow\infty} P_{H_1} \left[\left|\left|\sqrt{n} ({\dot{\mbox{\boldmath $\mu$}}}_{\gamma,n} - \mbox{\boldmath $\mu$}_{0})\right|\right|^2 > c_{\alpha}\right]$\\ $= \displaystyle\lim_{n\rightarrow\infty} P_{H_1} \left[\left|\left|\sqrt{n} ({\dot{\mbox{\boldmath $\mu$}}}_{\gamma,n} - \mbox{\boldmath $\mu$}_{1}+\mbox{\boldmath $\mu$}_{1}-\mbox{\boldmath $\mu$}_{0})\right|\right|^2 > c_{\alpha}\right]$\\$ = \displaystyle\lim_{n\rightarrow\infty} P_{H_1} \left[\left|\left|\sqrt{n} ({\dot{\mbox{\boldmath $\mu$}}}_{\gamma,n} - \mbox{\boldmath $\mu$}_{1})\right|\right|^2 + \left|\left|\sqrt{n}(\mbox{\boldmath $\mu$}_{1}-\mbox{\boldmath $\mu$}_{0})\right|\right|^2 + 2 \left<\sqrt{n} ({\dot{\mbox{\boldmath $\mu$}}}_{\gamma,n} - \mbox{\boldmath $\mu$}_{1}), \sqrt{n}(\mbox{\boldmath $\mu$}_{1}-\mbox{\boldmath $\mu$}_{0}) \right>  > c_{\alpha}\right]$ \\$=\displaystyle\lim_{n\rightarrow\infty} P_{H_1} \left[\left|\left|\sqrt{n} ({\dot{\mbox{\boldmath $\mu$}}}_{\gamma,n} - \mbox{\boldmath $\mu$}_{1})\right|\right|^2 > c_{\alpha} - \left|\left|\sqrt{n}(\mbox{\boldmath $\mu$}_{1}-\mbox{\boldmath $\mu$}_{0})\right|\right|^2 - 2n \left<({\dot{\mbox{\boldmath $\mu$}}}_{\gamma,n} - \mbox{\boldmath $\mu$}_{1}), (\mbox{\boldmath $\mu$}_{1}-\mbox{\boldmath $\mu$}_{0}) \right>\right]$ \\$= \displaystyle\lim_{n\rightarrow\infty}P_{H_1} \left[\left|\left|\sqrt{n} ({\dot{\mbox{\boldmath $\mu$}}}_{\gamma,n} - \mbox{\boldmath $\mu$}_{1})\right|\right|^2 > c_{\alpha} - n\left|\left|(\mbox{\boldmath $\mu$}_{1}-\mbox{\boldmath $\mu$}_{0})\right|\right|^2 \right]$ 
since under $H_{1}$, ${\dot{\mbox{\boldmath $\mu$}}}_{\gamma,n} \xrightarrow{a.s.} \mbox{\boldmath $\mu$}_{1}$ \\$ \rightarrow 1 $ as $n\rightarrow\infty$. 

The last implication follows from the fact that $\left|\left|\sqrt{n} ({\dot{\mbox{\boldmath $\mu$}}}_{\gamma,n} - \mbox{\boldmath $\mu$}_{1})\right|\right|^2$ converges weakly to  the distribution of $\sum\limits_{i = 1}^{d}\lambda_{i}Z_{i}^{2}$ under $H_{1}$, and $c_{\alpha} - n\left|\left|(\mbox{\boldmath $\mu$}_{1}-\mbox{\boldmath $\mu$}_{0})\right|\right|^2$ converges to $-\infty$ as $n\rightarrow\infty$. This fact leads to the result.

\vspace{0.1 in}

\noindent{\bf Proof of Proposition 2.1.:} For testing $H_{0}: \mbox{\boldmath $\mu$} = \mbox{\boldmath $\mu$}_0$ against $H_{1}: \mbox{\boldmath $\mu$}\neq \mbox{\boldmath $\mu$}_0$, the power of the test based on $T_{n}^{2}$ will be $P_{H_{1}}[T_{n}^{2} > c^{*}_{\alpha}]$, where $c^{*}_{\alpha}$ is the $(1 - \alpha)$-th $(0 < \alpha < 1)$ quantile of the distribution of $\sum\limits_{i = 1}^{d}\lambda^{*}_{i}Z_{i}^{*2}$. Here, $\lambda^{*}_{i}$s are the eigen values of $\sigma_{2}^{2}\Sigma$, and $Z^{*}_{i}$'s are i.i.d.\ $N(0, 1)$ random variables. In view of the orthogonal decomposition of multivariate normal distribution, $T_{n}^{2}$ converges weakly to the distribution of  $\sum\limits_{i = 1}^{d}\lambda_{i}^{*}Z_{i}^{*2}$, and hence, the asymptotic size of the test based on $T_{n}^{2}$ is $\alpha$. As earlier, let us again denote $\mbox{\boldmath $\mu$} = \mbox{\boldmath $\mu$}_{1} (\neq \mbox{\boldmath $\mu$}_0)$ under $H_{1}$, and we now have 

$\displaystyle\lim_{n\rightarrow\infty} P_{H_1} \left[T_{n}^{2} > c^{*}_{\alpha}\right]$ \\$= \displaystyle\lim_{n\rightarrow\infty} P_{H_1} \left[\left|\left|\sqrt{n} ({\hat{\mbox{\boldmath $\mu$}}}_{SM} - \mbox{\boldmath $\mu$}_{0})\right|\right|^2 > c^{*}_{\alpha}\right]$\\ $= \displaystyle\lim_{n\rightarrow\infty} P_{H_1} \left[\left|\left|\sqrt{n} ({\hat{\mbox{\boldmath $\mu$}}}_{SM} - \mbox{\boldmath $\mu$}_{1}+\mbox{\boldmath $\mu$}_{1}-\mbox{\boldmath $\mu$}_{0})\right|\right|^2 > c^{*}_{\alpha}\right]$\\$ = \displaystyle\lim_{n\rightarrow\infty} P_{H_1} \left[\left|\left|\sqrt{n} ({\hat{\mbox{\boldmath $\mu$}}}_{SM} - \mbox{\boldmath $\mu$}_{1})\right|\right|^2 + \left|\left|\sqrt{n}(\mbox{\boldmath $\mu$}_{1}-\mbox{\boldmath $\mu$}_{0})\right|\right|^2 + 2 \left<\sqrt{n} ({\hat{\mbox{\boldmath $\mu$}}}_{SM} - \mbox{\boldmath $\mu$}_{1}), \sqrt{n}(\mbox{\boldmath $\mu$}_{1}-\mbox{\boldmath $\mu$}_{0}) \right>  > c^{*}_{\alpha}\right]$ \\$=\displaystyle\lim_{n\rightarrow\infty} P_{H_1} \left[\left|\left|\sqrt{n} ({\hat{\mbox{\boldmath $\mu$}}}_{SM} - \mbox{\boldmath $\mu$}_{1})\right|\right|^2 > c^{*}_{\alpha} - \left|\left|\sqrt{n}(\mbox{\boldmath $\mu$}_{1}-\mbox{\boldmath $\mu$}_{0})\right|\right|^2 - 2n \left<({\hat{\mbox{\boldmath $\mu$}}}_{SM} - \mbox{\boldmath $\mu$}_{1}), (\mbox{\boldmath $\mu$}_{1}-\mbox{\boldmath $\mu$}_{0}) \right>\right]$ \\$= \displaystyle\lim_{n\rightarrow\infty}P_{H_1} \left[\left|\left|\sqrt{n} ({\hat{\mbox{\boldmath $\mu$}}}_{SM} - \mbox{\boldmath $\mu$}_{1})\right|\right|^2 > c^{*}_{\alpha} - n\left|\left|(\mbox{\boldmath $\mu$}_{1}-\mbox{\boldmath $\mu$}_{0})\right|\right|^2 \right]$ (since under $H_{1}$, ${\hat{\mbox{\boldmath $\mu$}}}_{SM} \xrightarrow{a.s.}\mbox{\boldmath $\mu$}_{1}$) \\$ \rightarrow 1 $ as $n\rightarrow\infty$. 

The last implication follows from the fact that $\left|\left|\sqrt{n} ({\hat{\mbox{\boldmath $\mu$}}}_{SM} - \mbox{\boldmath $\mu$}_{1})\right|\right|^2$ converges weakly to  the distribution of $\sum\limits_{i = 1}^{d}\lambda^{*}_{i}Z_{i}^{*2}$ under $H_{1}$, and $c^{*}_{\alpha} - n\left|\left|(\mbox{\boldmath $\mu$}_{1}-\mbox{\boldmath $\mu$}_{0})\right|\right|^2$ converges to $-\infty$ as $n\rightarrow\infty$. This completes the proof.

\vspace{0.1 in}

\noindent{\bf Proof of Proposition 2.2.:} In order to test $H_{0}: \mbox{\boldmath $\mu$} = \mbox{\boldmath $\mu$}_0$ against $H_{1}: \mbox{\boldmath $\mu$}\neq \mbox{\boldmath $\mu$}_0$, the form of the power function of the test based on $T_{n}^{3}$ is $P_{H_{1}}[T_{n}^{3} > c^{**}_{\alpha}]$, where $c^{**}_{\alpha}$ is same as described in the statement of the proposition. In view of the orthogonal decomposition of multivariate normal distribution, $T_{n}^{3}$ converges weakly to the distribution of  $\sum\limits_{i = 1}^{d}\lambda_{i}^{**}Z_{i}^{**2}$, and hence, the asymptotic size of the test based on $T_{n}^{3}$ is $\alpha$
Further, we denote that $\mbox{\boldmath $\mu$} = \mbox{\boldmath $\mu$}_{1} (\neq \mbox{\boldmath $\mu$}_0)$ under $H_{1}$, and we then have

$\displaystyle\lim_{n\rightarrow\infty} P_{H_1} \left[T_{n}^{3} > c^{**}_{\alpha}\right]$ \\$= \displaystyle\lim_{n\rightarrow\infty} P_{H_1} \left[\left|\left|\sqrt{n} ({\hat{\mbox{\boldmath $\mu$}}}_{CM} - \mbox{\boldmath $\mu$}_{0})\right|\right|^2 > c^{**}_{\alpha}\right]$\\ $= \displaystyle\lim_{n\rightarrow\infty} P_{H_1} \left[\left|\left|\sqrt{n} ({\hat{\mbox{\boldmath $\mu$}}}_{CM} - \mbox{\boldmath $\mu$}_{1}+\mbox{\boldmath $\mu$}_{1}-\mbox{\boldmath $\mu$}_{0})\right|\right|^2 > c^{**}_{\alpha}\right]$\\$ = \displaystyle\lim_{n\rightarrow\infty} P_{H_1} \left[\left|\left|\sqrt{n} ({\hat{\mbox{\boldmath $\mu$}}}_{CM} - \mbox{\boldmath $\mu$}_{1})\right|\right|^2 + \left|\left|\sqrt{n}(\mbox{\boldmath $\mu$}_{1}-\mbox{\boldmath $\mu$}_{0})\right|\right|^2 + 2 \left<\sqrt{n} ({\hat{\mbox{\boldmath $\mu$}}}_{CM} - \mbox{\boldmath $\mu$}_{1}), \sqrt{n}(\mbox{\boldmath $\mu$}_{1}-\mbox{\boldmath $\mu$}_{0}) \right>  > c^{**}_{\alpha}\right]$ \\$=\displaystyle\lim_{n\rightarrow\infty} P_{H_1} \left[\left|\left|\sqrt{n} ({\hat{\mbox{\boldmath $\mu$}}}_{CM} - \mbox{\boldmath $\mu$}_{1})\right|\right|^2 > c^{**}_{\alpha} - \left|\left|\sqrt{n}(\mbox{\boldmath $\mu$}_{1}-\mbox{\boldmath $\mu$}_{0})\right|\right|^2 - 2n \left<({\hat{\mbox{\boldmath $\mu$}}}_{CM} - \mbox{\boldmath $\mu$}_{1}), (\mbox{\boldmath $\mu$}_{1}-\mbox{\boldmath $\mu$}_{0}) \right>\right]$ \\$= \displaystyle\lim_{n\rightarrow\infty}P_{H_1} \left[\left|\left|\sqrt{n} ({\hat{\mbox{\boldmath $\mu$}}}_{CM} - \mbox{\boldmath $\mu$}_{1})\right|\right|^2 > c^{**}_{\alpha} - n\left|\left|(\mbox{\boldmath $\mu$}_{1}-\mbox{\boldmath $\mu$}_{0})\right|\right|^2 \right]$ (since under $H_{1}$, ${\hat{\mbox{\boldmath $\mu$}}}_{CM} \xrightarrow{a.s.} \mbox{\boldmath $\mu$}_{1}$) \\$ \rightarrow 1 $ as $n\rightarrow\infty$. 

The last implication follows from the fact that $\left|\left|\sqrt{n} ({\hat{\mbox{\boldmath $\mu$}}}_{CM} - \mbox{\boldmath $\mu$}_{1})\right|\right|^2$ converges weakly to the same distribution as described in the statement of the proposition, and $c^{**}_{\alpha} - n\left|\left|(\mbox{\boldmath $\mu$}_{1}-\mbox{\boldmath $\mu$}_{0})\right|\right|^2$ converges to $-\infty$ as $n\rightarrow\infty$. This completes the proof. 

\vspace{0.1 in}

\noindent{\bf Proof of Proposition 2.3.:} In order to test $H_{0}: \mbox{\boldmath $\mu$} = \mbox{\boldmath $\mu$}_0$ against $H_{1}: \mbox{\boldmath $\mu$}\neq \mbox{\boldmath $\mu$}_0$, the form of the power function of the test based on $T_{n}^{4}$ is $P_{H_{1}}[T_{n}^{4} > c^{***}_{\alpha}]$, where $c^{***}_{\alpha}$ is same as described in the statement of the proposition. In view of the orthogonal decomposition of multivariate normal distribution, $T_{n}^{4}$ converges weakly to the distribution of  $\sum\limits_{i = 1}^{d}\lambda_{i}^{***}Z_{i}^{***2}$, and hence, the asymptotic size of the test based on $T_{n}^{4}$ is $\alpha$
Further, we denote that $\mbox{\boldmath $\mu$} = \mbox{\boldmath $\mu$}_{1} (\neq \mbox{\boldmath $\mu$}_0)$ under $H_{1}$, and we then have

$\displaystyle\lim_{n\rightarrow\infty} P_{H_1} \left[T_{n}^{4} > c^{***}_{\alpha}\right]$ \\$= \displaystyle\lim_{n\rightarrow\infty} P_{H_1} \left[\left|\left|\sqrt{n} ({\hat{\mbox{\boldmath $\mu$}}}_{HL} - \mbox{\boldmath $\mu$}_{0})\right|\right|^2 > c^{***}_{\alpha}\right]$\\ $= \displaystyle\lim_{n\rightarrow\infty} P_{H_1} \left[\left|\left|\sqrt{n} ({\hat{\mbox{\boldmath $\mu$}}}_{HL} - \mbox{\boldmath $\mu$}_{1}+\mbox{\boldmath $\mu$}_{1}-\mbox{\boldmath $\mu$}_{0})\right|\right|^2 > c^{***}_{\alpha}\right]$\\$ = \displaystyle\lim_{n\rightarrow\infty} P_{H_1} \left[\left|\left|\sqrt{n} ({\hat{\mbox{\boldmath $\mu$}}}_{HL} - \mbox{\boldmath $\mu$}_{1})\right|\right|^2 + \left|\left|\sqrt{n}(\mbox{\boldmath $\mu$}_{1}-\mbox{\boldmath $\mu$}_{0})\right|\right|^2 + 2 \left<\sqrt{n} ({\hat{\mbox{\boldmath $\mu$}}}_{HL} - \mbox{\boldmath $\mu$}_{1}), \sqrt{n}(\mbox{\boldmath $\mu$}_{1}-\mbox{\boldmath $\mu$}_{0}) \right>  > c^{***}_{\alpha}\right]$ \\$=\displaystyle\lim_{n\rightarrow\infty} P_{H_1} \left[\left|\left|\sqrt{n} ({\hat{\mbox{\boldmath $\mu$}}}_{HL} - \mbox{\boldmath $\mu$}_{1})\right|\right|^2 > c^{***}_{\alpha} - \left|\left|\sqrt{n}(\mbox{\boldmath $\mu$}_{1}-\mbox{\boldmath $\mu$}_{0})\right|\right|^2 - 2n \left<({\hat{\mbox{\boldmath $\mu$}}}_{HL} - \mbox{\boldmath $\mu$}_{1}), (\mbox{\boldmath $\mu$}_{1}-\mbox{\boldmath $\mu$}_{0}) \right>\right]$ \\$= \displaystyle\lim_{n\rightarrow\infty}P_{H_1} \left[\left|\left|\sqrt{n} ({\hat{\mbox{\boldmath $\mu$}}}_{HL} - \mbox{\boldmath $\mu$}_{1})\right|\right|^2 > c^{***}_{\alpha} - n\left|\left|(\mbox{\boldmath $\mu$}_{1}-\mbox{\boldmath $\mu$}_{0})\right|\right|^2 \right]$ (since under $H_{1}$, ${\hat{\mbox{\boldmath $\mu$}}}_{HL} \xrightarrow{a.s.} \mbox{\boldmath $\mu$}_{1}$) \\$ \rightarrow 1 $ as $n\rightarrow\infty$. 

The last implication follows from the fact that $\left|\left|\sqrt{n} ({\hat{\mbox{\boldmath $\mu$}}}_{HL} - \mbox{\boldmath $\mu$}_{1})\right|\right|^2$ converges weakly to the same distribution as described in the statement of the proposition, and $c^{***}_{\alpha} - n\left|\left|(\mbox{\boldmath $\mu$}_{1}-\mbox{\boldmath $\mu$}_{0})\right|\right|^2$ converges to $-\infty$ as $n\rightarrow\infty$. This completes the proof.

\vspace{0.1 in}

\noindent {\bf Proof of Theorem 2.2.:} At first, one needs to know the form of $g(x)$ in the expressions of $e_{1}(d)$ and $e_{2}(d)$  that provided in the first paragraph in Section 2.2. Note that for a $d$-dimensional Gaussian distribution, $g(x) = e^{-\frac{x}{2}}$, in the case of a $d$-dimensional Cauchy distribution, $g(x) = \frac{1}{(1 + x)^{\frac{d + 1}{2}}}$, and for the given $d$-dimensional Spherical  distribution, $g(x)= e^{-x^{100}}$. Now, using the form of $g(x)$, we have $e_{1}(d) = \frac{\gamma}{{2\pi}^\frac{d}{2}}$, $e_{2}(d) = \frac{\gamma
{\pi}^{1-\frac{d}{2}}}{{2}^(1+\frac{d}{2})}$ and $e_3(d) = \frac{\gamma \pi^{1-\frac{d}{2}}}{3 \times {(2)}^\frac{d}{2}}$ for a $d$-dimensional Gaussian distribution. Since $\pi > 1$, one can conclude that $\displaystyle\lim_{d\rightarrow\infty} e_{1}(d) = 0$. Next, note that $e_{2}(d) = \frac{\gamma{\pi}^{1-\frac{d}{2}}}{{2}^(1+\frac{d}{2})} = \frac{\gamma\pi}{2\times (2\pi)^{\frac{d}{2}}}\rightarrow 0$ and $e_{3}(d) = \frac{\gamma{\pi}^{1-\frac{d}{2}}}{3 \times {(2)}^\frac{d}{2}} = \frac{\gamma\pi}{3\times (2\pi)^{\frac{d}{2}}}\rightarrow 0$ as $d\rightarrow\infty$ since $2\pi > 1$. 
Further, note that since $Var(Y_{1}) = \infty$ for a Cauchy distribution, we have $e_{1}(d) = \infty$ for all $d$, and hence, $\displaystyle\lim_{d\rightarrow\infty} e_{1}(d) = \infty$ for a $d$-dimensional Cauchy distribution. Similarly, using the form of $g(x)$ associated with Cauchy distribution, we have  $e_{2}(d) = \frac{\gamma d {\pi}^{\frac{3-d}{2}} \Gamma({\frac{d+1}{2}})}{4}$ and $e_{3}(d) = \frac{\gamma d {\pi}^{\frac{3-d}{2}} \Gamma({\frac{d+1}{2}})}{12}$. In order to investigate the limiting properties of $e_{2}(d)$, we consider $\frac{e_{2}(d + 1)}{e_{2}(d)} = \frac{d + 1}{d}\times\frac{1}{\sqrt{\pi}}\times\frac{\Gamma(d/2 + 1)}{\Gamma(d/2 + 1/2)}$. Next, using Stirling's approximation formula: $\Gamma(n + 1) = \sqrt{2\pi}e^{-n} n^{n + 1/2}$ when $n$ is an integer and $n\rightarrow\infty$, we have 

\vspace{-0.2in}

\begin{eqnarray*}
\lim_{d\rightarrow\infty}\frac{e_{2}(d + 1)}{e_{2}(d)} & =& \lim_{d\rightarrow\infty}\frac{d + 1}{d}\times\frac{1}{\sqrt{\pi}}\times\frac{\Gamma(\frac{d}{2} + 1)}{\Gamma(\frac{d}{2} + \frac{1}{2})}\\
& = & 1\times\frac{1}{\sqrt{\pi}} \times\lim_{d\rightarrow\infty}\frac{\Gamma(\frac{d}{2} + 1)}{\Gamma(\frac{d}{2} + \frac{1}{2})}\\
& = & 1\times\frac{1}{\sqrt{\pi}} \times\lim_{d\rightarrow\infty}\frac{\sqrt{2\pi} e^{-\frac{d}{2}} (\frac{d}{2})^{\frac{d}{2} + 1}}{\sqrt{2\pi}e^{-(\frac{d}{2} - \frac{1}{2})} (\frac{d}{2} - \frac{1}{2})^{\frac{d}{2} + \frac{1}{2}}}\\
& = & 1\times\frac{1}{\sqrt{\pi}}\times\frac{1}{\sqrt{2e}} \times\lim_{d\rightarrow\infty}\frac{d}{\sqrt{d - 1}}\left(1 + \frac{1}{d - 1}\right)^{\frac{d}{2}}\\
& = & 1\times\frac{1}{\sqrt{\pi}}\times\frac{1}{\sqrt{2e}}\times\lim_{d\rightarrow\infty}\frac{d}{\sqrt{d - 1}}\times\lim_{d\rightarrow\infty}\left(1 + \frac{1}{d - 1}\right)^{\frac{d}{2}}\\
&=&\infty~\mbox{(since $\displaystyle\lim_{d\rightarrow\infty}\frac{d}{\sqrt{d - 1}} = \infty$)}.
\end{eqnarray*}  

\noindent This implies that $\displaystyle\lim_{d\rightarrow\infty} e_{2}(d) = \infty$. Arguing in similar way one can say that $\displaystyle\lim_{d\rightarrow\infty} e_{3}(d) = \infty$. Furthermore, for $d$-dimensional Spherical  distribution with $g(x) = e^{-x^{100}}$, we have $e_{1} (d) = \frac{100 d\gamma \Gamma(\frac{d}{2})}{\pi^{\frac{d}{2}} \Gamma(\frac{1}{100}(\frac{d}{2}+1))}$, $e_{2}(d) = \frac{d\gamma \Gamma(\frac{d}{2}) (\Gamma(\frac{1}{200}))^2}{400 \pi^{\frac{d}{2}} \Gamma(\frac{1}{100}(\frac{d}{2}+1))}$ and $e_{3}(d) = \frac{53188.48 d\gamma \Gamma(\frac{d}{2})}{ \pi^{\frac{d}{2}} \Gamma(\frac{1}{100}(\frac{d}{2}+1))} $. In order to investigate the limiting properties of $e_{1}(d)$ the repeated use of Stirling's approximation formula leads to

\vspace{-0.2in}

\begin{eqnarray*}
\lim_{d\rightarrow\infty}\frac{e_{1}(d + 1)}{e_{1}(d)} & =& \lim_{d\rightarrow\infty}\frac{d + 1}{d}\times\frac{1}{\sqrt{\pi}}\times\frac{\Gamma(\frac{d+1}{2})}{\Gamma(\frac{d}{2})}\times \frac{\Gamma(\frac{1}{100}(\frac{d}{2}+1))}{\Gamma(\frac{1}{100}(\frac{d+1}{2}+1))}\\
& = & 1\times\frac{1}{\sqrt{\pi}} \times\lim_{d\rightarrow\infty}\frac{\Gamma(\frac{d}{2} + \frac{1}{2})}{\Gamma(\frac{d}{2})} \times \frac{\Gamma(\frac{d}{200}+\frac{1}{100})}{\Gamma(\frac{d+1}{200}+\frac{1}{100})}\\
& = & 1\times\frac{1}{\sqrt{\pi}} \times\lim_{d\rightarrow\infty}\frac{\sqrt{2\pi}e^{-(\frac{d}{2} - \frac{1}{2})} (\frac{d}{2} - \frac{1}{2})^{\frac{d}{2}}}{\sqrt{2\pi}e^{-(\frac{d}{2} - 1)} (\frac{d}{2} - 1)^{\frac{d}{2} - \frac{1}{2}}} \times \frac{\sqrt{2\pi}e^{-(\frac{d}{200} - \frac{198}{200})} (\frac{d}{200} - \frac{198}{200})^{\frac{d}{200} - \frac{98}{200}}}{\sqrt{2\pi}e^{-(\frac{d}{200} - \frac{197}{200})} (\frac{d}{200} - \frac{197}{200})^{\frac{d}{200} - \frac{97}{200}}} \\
& = & 1\times\frac{1}{\sqrt{\pi}}\times\frac{(200e)^{\frac{1}{200}}}{\sqrt{2e}} \times\lim_{d\rightarrow\infty}{\sqrt{d - 2}}\left(1 - \frac{1}{d - 2}\right)^{\frac{d}{2}} \times \lim_{d\rightarrow\infty}{\frac{\left(1 - \frac{1}{d - 197}\right)^{\frac{d}{200}-\frac{98}{200}}}{(d-197)^{\frac{1}{200}}}}\\
& = & 1\times\frac{1}{\sqrt{\pi}}\times\frac{1}{\sqrt{2e}}\times\lim_{d\rightarrow\infty}\frac{\sqrt{d - 2}}{(d-197)^{\frac{1}{200}}}\times\lim_{d\rightarrow\infty}\left(1 - \frac{1}{d - 2}\right)^{\frac{d}{2}} \times \left(1 - \frac{1}{d - 197}\right)^{\frac{d}{200}-\frac{98}{200}}\\
&=&\infty~\mbox{(since $\displaystyle\lim_{d\rightarrow\infty}\frac{\sqrt{d - 2}}{(d-197)^{\frac{1}{200}}} = \infty$)}.
\end{eqnarray*}  

This leads us to $\displaystyle\lim_{d\rightarrow\infty} e_{1}(d) = \infty$. Similarly one can say that $\displaystyle\lim_{d\rightarrow\infty} e_{2}(d) = \infty$ and $\displaystyle\lim_{d\rightarrow\infty} e_{3}(d) = \infty$, and hence the proof is complete. 

\vspace{0.1 in}

\noindent \textbf{Proof of Theorem 3.1.:} In order to establish the contiguity of the distributions associated with sequence $\{H_{1n}\}$ relative to those of $\{H_{0n}\}$, it is enough to show that $\Lambda_n$, the logarithm of the likelihood ratio, converges weakly to a random variable associated with a normal distribution with location parameter $= -\frac{\sigma^{2}}{2}$ and variance $= \sigma^{2}$, where $\sigma$ is a positive constant (see Hajek, Sidak and Sen (1999), p. 254, Corollary to Lecam's first lemma). Let ${\bf y}_{1}$, $\ldots$, ${\bf y}_{n}$ be i.i.d.\ random variables with the probability density function $f(y; (.), \Sigma)$, where $(.)$ denotes the location parameter involved in the distribution, and $\Sigma$ is the scatter matrix. We now consider\\ 
\begin{eqnarray*}
\Lambda_n & = & \displaystyle\log\prod_{i = 1}^{n}\frac{f\left(\textbf{y}_i,\mbox{\boldmath $\mu$}_0+\frac{\mbox{\boldmath $\delta$}}{\sqrt{n}}\right)}{f\left(\textbf{y}_i,\mbox{\boldmath $\mu$}_0\right)} = \sum \limits_{i=1}^{n} \left\{\log f\left(\textbf{y}_i,\mbox{\boldmath $\mu$}_0+\frac{\mbox{\boldmath $\delta$}}{\sqrt{n}}\right) - 
 \log f\left(\textbf{y}_i,\mbox{\boldmath $\mu$}_0\right)\right\}\\
 & = & \sum\limits_{i=1}^{n}\left\{h\left(\textbf{y}_i,\mbox{\boldmath $\mu$}_0+\frac{\mbox{\boldmath $\delta$}}{\sqrt{n}}\right) - h\left(\textbf{y}_i,\mbox{\boldmath $\mu$}_0\right)\right\}~\mbox{(denoted as $h (.) = \log f(.)$)}\\
 & = & \sum\limits_{i=1}^{n} \left[ \frac{\mbox{\boldmath $\delta$}^{T}}{\sqrt{n}}\bigtriangledown\{h\left(\textbf{y}_i,\mbox{\boldmath $\mu$}_0\right)\} + \frac{1}{2n}\mbox{\boldmath $\delta$}^{T}H\{h\left(\textbf{y}_i,\mbox{\boldmath $\zeta$}_{n}\right)\}\mbox{\boldmath $\delta$}\right],
\end{eqnarray*}
where \mbox{\boldmath $\zeta$}$_{n}$ is any point lying on the straight line joining \mbox{\boldmath $\mu$}$_0$ and \mbox{\boldmath $\mu$}$_0+\frac{\mbox{\boldmath $\delta$}}{\sqrt{n}}$, $\bigtriangledown (.)$ denotes the gradient vector of $(.)$, and $H(.)$ denotes the Hessian matrix of $(.)$. Note that \mbox{\boldmath $\zeta$}$_{n}\rightarrow$ \mbox{\boldmath $\mu$}$_{0}$ as $n\rightarrow\infty$. 

It now follows from the central limit theorem that $\sum\limits_{i = 1}^{n}\frac{\mbox{\boldmath $\delta$}^{T}}{\sqrt{n}}\bigtriangledown\{h\left(\textbf{y}_i,\mbox{\boldmath $\mu$}_0\right)\}$ converges weakly to a random variable associated with normal distribution with mean $= E [\mbox{\boldmath $\delta$}^{T}\bigtriangledown\{h\left(\textbf{y}, \mbox{\boldmath $\mu$}_0\right)\}] = 0$ and variance $= E [\mbox{\boldmath $\delta$}^{T}\bigtriangledown\{h\left(\textbf{y}, \mbox{\boldmath $\mu$}_0\right)\}]^{2}$ when $ E [\mbox{\boldmath $\delta$}^{T}\bigtriangledown\{h\left(\textbf{y}, \mbox{\boldmath $\mu$}_0\right)\}]^{2} < \infty$ for all \mbox{\boldmath $\delta$}. Moreover, a direct algebra implies that $E [\mbox{\boldmath $\delta$}^{T}\bigtriangledown\{h\left(\textbf{y}, \mbox{\boldmath $\mu$}_0\right)\}]^{2} = - E [\mbox{\boldmath $\delta$}^{T}H\{h\left(\textbf{y}_i,\mbox{\boldmath $\mu$}_{0}\right)\}\mbox{\boldmath $\delta$}]$ (see, e.g., Shao (2003)). Hence, the asymptotic normality of $\sum\limits_{i = 1}^{n}\frac{\mbox{\boldmath $\delta$}^{T}}{\sqrt{n}}\bigtriangledown\{h\left(\textbf{y}_i,\mbox{\boldmath $\mu$}_0\right)\}$ holds when $E\left\{\frac{\partial^{2}}{\partial\mu_{i}\partial\mu_{j}}\log f({\bf y},  \mbox{\boldmath $\mu$}) \right\} < \infty$, where $\mu_{i}$ and $\mu_{j}$ are the $i$-th and the $j$-th components of \mbox{\boldmath $\mu$}.

Next, the other term $\frac{1}{2n}\sum\limits_{i = 1}^{n}\mbox{\boldmath $\delta$}^{T}H\{h\left(\textbf{y}_i,\mbox{\boldmath $\zeta$}_{n}\right)\}\mbox{\boldmath $\delta$}\stackrel{p}\rightarrow\frac{1}{2}E [\mbox{\boldmath $\delta$}^{T}H\{h\left(\textbf{y},\mbox{\boldmath $\mu$}_{0}\right)\}\mbox{\boldmath $\delta$}]$ in view of weak law of large number and \mbox{\boldmath $\zeta$}$_{n}\rightarrow$ \mbox{\boldmath $\mu$}$_{0}$ as $n\rightarrow\infty$. Finally, using Slutsky's result, one can conclude that $\Lambda_n$ converges weakly to a normal distribution with mean $= \frac{1}{2}E [\mbox{\boldmath $\delta$}^{T}H\{h\left(\textbf{y}_i,\mbox{\boldmath $\mu$}_{0}\right)\}\mbox{\boldmath $\delta$}]$ and variance $ = -  E [\mbox{\boldmath $\delta$}^{T}H\{h\left(\textbf{y}_i,\mbox{\boldmath $\mu$}_{0}\right)\}\mbox{\boldmath $\delta$}]$, where $E [\mbox{\boldmath $\delta$}^{T}H\{h\left(\textbf{y}_i,\mbox{\boldmath $\mu$}_{0}\right)\}\mbox{\boldmath $\delta$}]$ is a negative constant. Hence, the proof is complete. $\hfill\Box$

\vspace{0.1 in}

\noindent \textbf{Proof of Theorem 3.2.:} To establish this result, one first needs to show that the joint distribution of $\{\sqrt{n} ({\dot{\mbox{\boldmath $\mu$}}}_{\gamma,n} - \mbox{\boldmath $\mu$}_0), \Lambda_{n}\}$ is asymptotically Gaussian under $H_{0}$, which is asserted in Le Cam's third lemma (see, e.g., Hajek, Sidak and Sen (1999)). Note that Lemma 1 asserts, $\sqrt{n} ({\dot{\mbox{\boldmath $\mu$}}}_{\gamma,n} - \mbox{\boldmath $\mu$}_0)$ converges weakly to a $d$-dimensional Gaussian distribution under $H_{0}$, and further, it follows from the proof of Theorem 3.1. that $\Lambda_{n}$ converges weakly to a univariate Gaussian distribution with certain parameters under some conditions. Suppose now that ${\bf L}\in\mathbb{R}^{d}$ and $m\in\mathbb{R}$ are two arbitrary constants. In view of the linearization of  $\sqrt{n} ({\dot{\mbox{\boldmath $\mu$}}}_{\gamma,n} - \mbox{\boldmath $\mu$}_0)$ and $\Lambda_{n}$ along with a direct application of central limit theorem, one can establish that ${\bf L}^{T}.\{\sqrt{n} ({\dot{\mbox{\boldmath $\mu$}}}_{\gamma,n} - \mbox{\boldmath $\mu$}_0)\} + m\Lambda_{n}$ converges weakly to a univariate Gaussian distribution under $H_{0}$, where $(.)$ denotes the inner product. This fact implies that the joint distribution of $\{\sqrt{n} ({\dot{\mbox{\boldmath $\mu$}}}_{\gamma,n} - \mbox{\boldmath $\mu$}_0), \Lambda_{n}\}$ is asymptotically $(d + 1)$-dimensional Gaussian under $H_{0}$. Note that the $i$-th component of $d$-dimensional covariance between $\sqrt{n} ({\dot{\mbox{\boldmath $\mu$}}}_{\gamma,n} - \mbox{\boldmath $\mu$}_0)$ and $\Lambda_{n}$ is $E\left\{({\dot{\mbox{\boldmath $\mu$}}}_{\gamma,n, i} - \mbox{\boldmath $\mu$}_{0, i})\sum\limits_{j = 1}^{d}\delta_{j}\frac{\partial g({\bf y}, \mbox{\boldmath $\mu$})}{\partial \mu_{j}}|_{\mbox{\boldmath $\mu$} = \mbox{\boldmath $\mu$}_{0}}\right\}$, where $\delta_{i}$, ${\dot{\mbox{\boldmath $\mu$}}}_{\gamma, n, i}$, \mbox{\boldmath $\mu$}$_{0, i}$ and $\mu_{i}$ are the $i$-th component of \mbox{\boldmath $\delta$}, ${\dot{\mbox{\boldmath $\mu$}}}_{\gamma, n}$, \mbox{\boldmath $\mu$}$_{0}$ and \mbox{\boldmath $\mu$}, respectively, and $g({\bf y}, \mbox{\boldmath $\mu$}) = \log f({\bf y}, \mbox{\boldmath $\mu$})$.
 
Now, using Le Cam's third lemma, one can directly establish that under $H_{n}$,  $\{\sqrt{n} ({\dot{\mbox{\boldmath $\mu$}}}_{\gamma,n} - \mbox{\boldmath $\mu$}_0)\}$ converges weakly to $d$-dimensional Gaussian distribution with mean vector ${\bf a} = (a_{1}, \ldots, a_{d})$ and variance-covariance matrix $= \left[\frac{1}{d\gamma} \frac{\pi^\frac{d}{2}}{\Gamma{(\frac{d}{2}})}\int \limits_{0}^\infty x^{\frac{d}{2}} g(x)dx\right]\Sigma$, where $a _{i} = \\E\left\{({\dot{\mbox{\boldmath $\mu$}}}_{\gamma,n, i} - \mbox{\boldmath $\mu$}_{0, i})\sum\limits_{j = 1}^{d}\delta_{j}\frac{\partial g({\bf y}, \mbox{\boldmath $\mu$})}{\partial \mu_{j}}|_{\mbox{\boldmath $\mu$} = \mbox{\boldmath $\mu$}_{0}}\right\}$. Hence, under $H_{n}$, for any orthogonal matrix $A$, 
$\{A\sqrt{n} ({\dot{\mbox{\boldmath $\mu$}}}_{\gamma,n} - \mbox{\boldmath $\mu$}_0 - {\bf a})\}$ converges weakly to a $d$-dimensional Gaussian distribution with mean vector $= {\bf 0}$ and variance-covariance matrix $= \text{Diag} (\lambda_{1}, \ldots, \lambda_{d})$, where $\lambda_{i}$ is the $i$-th eigenvalue of $\left[\frac{1}{d\gamma} \frac{\pi^\frac{d}{2}}{\Gamma{(\frac{d}{2}})}\int \limits_{0}^\infty x^{\frac{d}{2}} g(x)dx\right]\Sigma$, and consequently, as $A$ is an orthogonal matrix, i.e., $A^{T}A = A A^{T} = I_{d}$, $\{A\sqrt{n} ({\dot{\mbox{\boldmath $\mu$}}}_{\gamma,n} - \mbox{\boldmath $\mu$}_0 - {\bf a})\}^{T}\{A\sqrt{n} ({\dot{\mbox{\boldmath $\mu$}}}_{\gamma,n} - \mbox{\boldmath $\mu$}_0 - {\bf a})\} = n||{\dot{\mbox{\boldmath $\mu$}}}_{\gamma,n} - \mbox{\boldmath $\mu$}_0 - {\bf a}||^{2}$ converges weakly to the distribution of $\sum\limits_{i = 1}^{d}\lambda_{i}Z_{i}^{2}$, where $Z_{i}$'s are i.i.d.\ standard normal random variables. This fact directly imply that under $H_{n}$ and condition assumed in Theorem 3.1., $T_{n}^{1} = n||{\dot{\mbox{\boldmath $\mu$}}}_{\gamma,n} - \mbox{\boldmath $\mu$}_0||^{2}$ converges weakly to the distribution of $\sum\limits_{i = 1}^{d}\lambda_{i}(Z_{i} + a_{i})^{2}$.

Similarly, based on the Bahadur expansion of the median (see, e.g., Serfling (1980)), the Hodges-Lehmann estimator (see, e.g., Lehmann (2006)) and the definition of the sample mean vector, one can conclude that  under $H_{0}$, ${\bf L}^{T}.\{\sqrt{n} ({\hat{\mbox{\boldmath $\mu$}}}_{CM} - \mbox{\boldmath $\mu$}_0)\} + m\Lambda_{n}$, ${\bf L}^{T}.\{\sqrt{n} ({\hat{\mbox{\boldmath $\mu$}}}_{HL} - \mbox{\boldmath $\mu$}_0)\} + m\Lambda_{n}$ and ${\bf L}^{T}.\{\sqrt{n} ({\hat{\mbox{\boldmath $\mu$}}}_{SM} - \mbox{\boldmath $\mu$}_0)\} + m\Lambda_{n}$ converge weakly to a Gaussian random variable with certain parameters for arbitrary constants ${\bf L}\in\mathbb{R}^{d}$ and $m\in\mathbb{R}$. This fact along with Le Cam's third lemma imply that under $H_{n}$, both $\sqrt{n} ({\hat{\mbox{\boldmath $\mu$}}}_{CM} - \mbox{\boldmath $\mu$}_0)$, $\sqrt{n} ({\hat{\mbox{\boldmath $\mu$}}}_{HL} - \mbox{\boldmath $\mu$}_0)$ and $\sqrt{n} ({\hat{\mbox{\boldmath $\mu$}}}_{SM} - \mbox{\boldmath $\mu$}_0)$ converge weakly to a $d$-dimensional Gaussian random vector with non-zero mean vectors and certain variance-covariance matrix. The $i$-th component of the mean vector for $\sqrt{n} ({\hat{\mbox{\boldmath $\mu$}}}_{SM} - \mbox{\boldmath $\mu$}_0)$, $\sqrt{n} ({\hat{\mbox{\boldmath $\mu$}}}_{CM} - \mbox{\boldmath $\mu$}_0)$ and $\sqrt{n} ({\hat{\mbox{\boldmath $\mu$}}}_{HL} - \mbox{\boldmath $\mu$}_0)$ are $a_{i}^{*}$, $a_{i}^{**}$ and $a_{i}^{***}$, respectively, where $i = 1, \ldots, d$. Finally, as we argued earlier, based on the well-known orthogonalization of the multivariate Gaussian distribution, one can conclude that under $H_{n}$, $T_{n}^{2}$ converges weakly to $\sum\limits_{i = 1}^{d}\lambda^{*}_{i}(Z_{i}^{*} + a_{i}^{*})^{2}$, $T_{n}^{3}$ converges weakly to $\sum\limits_{i = 1}^{d}\lambda^{**}_{i}(Z_{i}^{**} + a_{i}^{**})^{2}$, and $T_{n}^{4}$ converges weakly to $\sum\limits_{i = 1}^{d}\lambda^{***}_{i}(Z_{i}^{***} + a_{i}^{***})^{2}$. The description of $\lambda_{i}^{*}$, $a_{i}^{*}$, $Z_{i}^{*}$, $\lambda_{i}^{**}$, $a_{i}^{**}$, $Z_{i}^{**}$, $\lambda_{i}^{***}$, $a_{i}^{***}$ and $Z_{i}^{***}$ are provided in the statement of the theorem. The proof is complete now. 

\vspace{0.1 in}

\noindent {\bf Proof of Corollary 3.1.:}  It follows from the assertion in Theorem 3.2. that under $H_{n}$, the power of the test based on $T_{n}^{1}$ is $P_{\mbox{\boldmath $\delta$}}\left[\sum\limits_{i = 1}^{d}\lambda_{i}(Z_{i} + a_{i})^{2} > c_{\alpha}\right]$. Here $c_{\alpha}$ is such that $\displaystyle P_{\mbox{\boldmath $\delta$ = {\bf 0}}}\left[\sum\limits_{i = 1}^{d}\lambda_{i}(Z_{i} + a_{i})^{2} > c_{\alpha}\right] = \alpha$ since $H_{n}$ coincides with $H_{0}$ when \mbox{\boldmath $\delta$}$\hspace{0.15 cm}= {\bf 0}$. Arguing exactly in a similar way, one can establish the asymptotic power of the tests based on $T_{n}^{2}$, $T_{n}^{3}$ and $T_n^4$ under $H_{n}$.

\section{Appendix B: Properties of Forward Search Estimator}

Here, we study a few fundamental properties of the multivariate forward search location estimator along with its performances.

\vspace{0.1 in}

\noindent {\bf (Property 1) Robustness of ${\dot{\mbox{\boldmath $\mu$}}}_{\gamma,n}$ :} 
The robustness property of ${\dot{\mbox{\boldmath $\mu$}}}_{\gamma,n}$ is described by the finite sample breakdown point, which is defined as follows. For the estimator ${\dot{\mbox{\boldmath $\mu$}}}_{\gamma,n}$ based on the data ${\cal{Y}} = \{{\bf y}_{1}, \ldots, {\bf y}_{n}\}$, its finite sample breakdown point is defined as $\epsilon({\dot{\mbox{\boldmath $\mu$}}}_{\gamma,n}, {\cal{Y}}) = \displaystyle \min_{m^\star\leq n^\star\leq n}\left\{\frac{n^\star}{n}:\sup_{{\cal{Y^{*}}}}\left|\left|{\dot{\mbox{\boldmath $\mu$}}}_{\gamma,n} - {\dot{\mbox{\boldmath $\mu$}}}_{\gamma,n}^{(n^\star)}\right|\right| = \infty\right\}, $ where $m^{*}$ is the cardinality of the initial subset, and ${\dot{\mbox{\boldmath $\mu$}}}_{\gamma,n}^{(n^\star)}$ is the forward search estimator of $\mbox{\boldmath $\mu$}$ based on a modified sample ${\cal{Y^{*}}} = \{{\bf y}^{*}_{1}, \ldots, {\bf y}^{*}_{n^{*}}, {\bf y}_{n^{*} + 1}, \ldots, {\bf y}_{n}\}$. 
The following theorem describes the breakdown point of ${\dot{\mbox{\boldmath $\mu$}}}_{\gamma,n}$.  

\vspace{0.1 in}

\noindent\textbf{Theorem 6.1.}
{\it Suppose that {\bf y}'s are in general position i.e., one cannot draw a hyperplane passing through all the observations, in the case of ${\dot{\mbox{\boldmath $\mu$}}}_{\gamma, n}$ for $\gamma$-th step. Then, we have $\epsilon({\dot{\mbox{\boldmath $\mu$}}}_{\gamma,n}, {\cal{Y}}) = 1-\gamma.$} 

\vspace{0.1 in}

\noindent {\bf Proof of Theorem 6.1.:} Suppose that the original observations are denoted by ${\cal{Y}} = \{{\bf y}_{1}, \ldots, {\bf y}_{n}\}$, and without loss of generality, first $n^{*} < n$ observations are corrupted. Let ${\cal{Y}}^{*} = \{{\bf y}_{1}^{*}, \ldots, {\bf y}_{n^{*}}^{*}, {\bf y}_{n^{*} + 1}, \ldots, {\bf y}_{n}\}$ denote the contaminated sample, where ${\bf y}_{i}^{*}$ are corrupted observations, $i = 1, \ldots, n^{*}$. It follows from the construction of the estimator that $\displaystyle\sup_{{\cal{Y}}^{*}}\left|\left|{\dot{\mbox{\boldmath $\mu$}}}_{\gamma,n} - {\dot{\mbox{\boldmath $\mu$}}}_{\gamma,n}^{(n^\star)}\right|\right| = \infty$ if and only if $\left|\left|\textbf{y}_{i^\star}^\star\right|\right|=\infty$ for any $i = 1, \ldots, n^{*}$. Without loss of generality, suppose that the aforementioned equivalent relationship holds for some $k$, and we than have 
$\eta_{k,\gamma,n} = 1$ since $\sum\limits_{i = 1}^{n}\eta_{i,\gamma, n} = m>0$. This fact implies that $Md^2_{k,n} = \infty$ for those choices of $k$. Let us further consider that there are $n_{1}^{*} < n^{*}$ many choices of $k$ for which $Md^2_{k,n} = \infty$. 
Now, in view of the definition of $\eta_{k,\gamma, n}$, we have $\eta_{j,\gamma,n} = I(Md^2_{j,n} \leq \delta^2_{\gamma, n}) = I(Md^2_{j, n} \leq Md^2_{(m), n}) = 0$ 
for $j = 1, \ldots, n_{1}^{*}$ when $n_{1}^{*} < n - m$. Further, note that $Md^2_{k,n} = \infty$ for any $k = 1, \ldots, n^{*}$ when $n_{1}^{*} = n^{*}$. Hence, $\epsilon({\dot{\mbox{\boldmath $\mu$}}}_{\gamma,n}, {\cal{Y}}) = 1-\gamma,$ and the proof is complete. 

\vspace{0.1 in}

\noindent\textbf{Remark 6.1.}
We would like to discuss on the breakdown point of ${\dot{\mbox{\boldmath $\mu$}}}_{\gamma,n}$.\ Note that $(1 - \gamma)$ is essentially the trimming proportion of the observations in the forward search estimator, and for that reason, it is expected that this estimator cannot be {\it breaking down} even in the presence of $(1 - \gamma)$ proportion outliers in the data.\ It indicates that $\gamma$ controls the breakdown point of ${\dot{\mbox{\boldmath $\mu$}}}_{\gamma,n}$.\ For instance, the breakdown point of ${\dot{\mbox{\boldmath $\mu$}}}_{\gamma,n}$ will achieve the highest possible value $= 1/2$ when $\gamma = 1/2$, and on the other hand, when $\gamma =1\Leftrightarrow m = n$, the breakdown point of the forward search estimator will be $ = 0$.\ Overall, the robustness behaviour of the forward search estimator is similar to any other trimming based estimator, e.g., the trimmed mean (see, e.g., Tukey (1948), Bickel (1965)).

\vspace{0.2 in}

\noindent {\bf (Property 2) Finite Sample Efficiency of ${\dot{\mbox{\boldmath $\mu$}}}_{\gamma,n}$ :} 

We here investigate the finite sample efficiency of ${\dot{\mbox{\boldmath $\mu$}}}_{\gamma,n}$ relative to the sample mean, the co-ordinate wise median and the Hodges-Lehmann estimator for $d$-dimensional  Gaussian distribution, $d$-dimensional Cauchy distribution with probability  density function $f_C({\bf x}) = (\Gamma((d + 1)/2)/{\pi}^{d/2}\Gamma(1/2))(1 + {\bf x}^{T}{\bf x})^{-(d +1)/2}$ and $d$-dimensional Spherical  distribution with  with $g(x) = e^{-x^{100}}$.\ In this simulation study, the finite sample efficiency of an estimator ${T}_{n}$ relative to ${T}_{n}^{'}$  is defined as $\{|COV(T_{n}^{'})|/|COV(T_{n})|\}^{1/d}$, where $|COV({T}_{n})|$ is the determinant of $COV({T}_{n}) =
\frac{1}{m}\sum\limits_{i = 1}^{m}({T}_{n}^{i} -\bar{{T}}_{n})({T}_{n}^{i} -\bar{{T}}_{n})^{T}$, and $m$ is
the number of Monte-Carlo replications. Here, 
${T}_{n}^{i}$ is the estimate of ${T}_{n}$ based on the $i$-th
replication, and $\bar{{T}}_{n} =  \frac{1}{m}\sum\limits_{i =
1}^{m}{T}_{n}^{i}$. In this simulation study, we consider $m = 1000$, and $n = 10$ and $100$, and the  results are summarized in Table 3.

Here, ${\dot{\mbox{\boldmath $\mu$}}}_{\gamma,n}$ performs well in term of finite sample efficiency for Cauchy distribution. The Hodges-Lehmann estimator also has better efficiency than the co-ordinate wise median for like Cauchy distribution. For normal distribution, the sample mean performs best although it fails to perform well for Cauchy distribution. The Hodges-Lehmann estimator is nearly as efficient as the sample mean for the normal distribution. For Spherical  distribution with  with $g(x) = e^{-x^{100}}$,  ${\dot{\mbox{\boldmath $\mu$}}}_{\gamma,n}$  performs best compared to other three estimators.

\vspace{0.1in}

\noindent {\bf Table 3:} Finite sample efficiencies of the multivariate forward search location estimator relative to the sample mean, the co-ordinate wise median and the Hodges-Lehmann estimator for different values of $n$, $d$ and various distributions.\ Here $\gamma = 1/2$.

\begin{center}
\begin{adjustbox}{max width=\textwidth}
\begin{tabular}{ |c|c|c|c|c|c|c|c|c|} 
\hline
 Normal distribution($n = 10$)   & $d = 2$ & $d = 4$ & $d = 10$ & $d = 20$ & $d = 50$ & $d = 100$\\
\hline
Sample mean  & 0.45 & 0.9 & 0.63 & 0.54 & 0.49 & 0.46 \\ 
\hline
CW median & 1.94 & 1.22 & 0.88 & 0.75 & 0.67 & 0.63 \\
\hline
HL estimator & 0.47 & 0.94 & 0.66 & 0.57 & 0.51 & 0.48\\ 
\hline
 Normal distribution($n = 100$) & $d = 2$ & $d = 4$ & $d = 10$ & $d = 20$ & $d = 50$ & $d = 100$ \\
\hline
Sample mean  & 0.52 & 0.98 & 0.75 & 0.65 & 0.58 & 0.55 \\ 
\hline
CW median & 2.43& 1.52 & 1.16& 1.004& 0.89 & 0.86\\
\hline
HL estimator & 0.54 & 1.03 & 0.79 & 0.68 & 0.6 & 0.58\\
\hline
Cauchy distribution($n = 10$) & $d = 2$ & $d = 4$ & $d = 10$ & $d = 20$ & $d = 50$ & $d = 100$\\
\hline
Sample mean  & 4.65 $\times 10^{4}$ & 8.06 $\times 10^{3}$ & 1.02 $\times 10^{3}$& 556.74 & 334.13 & 234.82\\ 
\hline
CW median &  2.51 & 2.07 & 1.71 & 1.64 & 1.56 & 1.53 \\
\hline
HL estimator & 1.83 & 1.69 & 0.98 & 0.97 & 0.92 & 0.9\\
\hline
Cauchy distribution($n = 100$) & $d = 2$ & $d = 4$ & $d = 10$ & $d = 20$ & $d = 50$ & $d = 100$\\
\hline
Sample mean  & 3.48 $\times 10^{4}$ & 1.8 $\times 10^{4}$ & 1.7 $\times 10^{4}$& 1.3 $\times 10^{4}$ & 3.9 $\times 10^{3}$& $5.5\times 10^{5}$ \\ 
\hline
CW median & 2.44 & 1.87 & 1.57 & 1.49 & 1.43 & 1.38 \\
\hline
HL estimator & 1.44 & 1.1 & 0.92 & 0.88 & 0.84 & 0.81\\
\hline
Spherical  distribution  with $g(x) = e^{-x^{100}}$ ($n = 10$) & $d = 2$ & $d = 4$ & $d = 10$ & $d = 20$ & $d = 50$ & $d = 100$\\
\hline
Sample mean  & 2.49 & 2.12 & 1.85 & 1.77 & 1.51 & 1.16\\ 
\hline
CW median & 2.48 & 2.1 & 1.84 & 1.75 & 1.5 & 1.13 \\
\hline
HL estimator & 2.39 & 1.92& 1.57 & 1.42 & 1.2 & 1.01\\ 
\hline
Spherical  distribution  with $g(x) = e^{-x^{100}}$ ($n = 100$) & $d = 2$ & $d = 4$ & $d = 10$ & $d = 20$ & $d = 50$ & $d = 100$\\
\hline
Sample mean  & 2.61 & 2.42 & 2.04 & 1.88 & 1.75 & 1.28\\ 
\hline
CW median & 2.5 & 2.37& 1.96 & 1.57 & 1.19 & 1.09 \\
\hline
HL estimator & 2.14 & 1.76 & 1.22 & 1.3 & 1.12 & 1.04\\ 
\hline
\end{tabular}
\end{adjustbox}
\end{center}


\vspace{0.2 in}

\noindent {\bf (Property 3) Asymptotic Efficiency of ${\dot{\mbox{\boldmath $\mu$}}}_{\gamma,n}$ :} In this study, we consider $d$-dimensional Gaussian, $d$-dimensional Cauchy distributions and $d$-dimensional Spherical  distribution with $g(x)=e^{-x^{100}}$ to carry out this study.\ The asymptotic efficiencies of ${\dot{\mbox{\boldmath $\mu$}}}_{\gamma,n}$ relative to the sample mean, the co-ordinate wise median and the Hodges-Lehmann estimator for the aforementioned distributions are reported in Table 4.

\vspace{0.1 in}


\noindent {\bf Table 4:} Asymptotic efficiencies of the multivariate forward search location estimator relative to the sample mean, the co-ordinate wise median and the Hodges-Lehmann estimator for different values of dimensions.\ Here $\gamma = 1/2$.

\vspace{0.10 in}
\begin{center}
\begin{adjustbox}{max width=\textwidth}  
\begin{tabular}{ |c|c|c|c|c|c|c|} 
\hline
 Normal distribution & $d = 2$ & $d = 4$ & $d = 10$ & $d = 20$ & $d = 50$ & $d = 100$\\
\hline
Sample mean  & 0.28 & 0.34 & 0.37 & 0.38& 0.39 & 0.4\\ 
\hline
CW median & 0.35 & 0.38 & 0.39 & 0.39& 0.4 & 0.4\\
\hline
HL estimator & 0.29 & 0.36 & 0.39 & 0.4 & 0.41 & 0.42\\ 
\hline
Cauchy distribution & $d = 2$ & $d = 4$ & $d = 10$ & $d = 20$ & $d = 50$ & $d = 100$ \\
\hline
Sample mean  &$\infty$ &$\infty$ & $\infty$&$\infty$& $\infty$& $\infty$\\ 
\hline
CW median & 0.35 & 0.59 & 0.91 & 1.22 & 1.83 & 2.52\\
\hline
HL estimator & 0.2 & 0.35 & 0.54 & 0.72 & 1.08 & 1.49\\ 
\hline
Spherical  distribution  with $g(x) = e^{-x^{100}}$ & $d = 2$ & $d = 4$ & $d = 10$ & $d = 20$ & $d = 50$ & $d = 100$ \\
\hline
Sample mean  &1.27 &1.49 & 1.85 &2.09 & 2.54 & 2.68\\ 
\hline
CW median & 1.14 &1.31 & 1.73 & 1.94 &2.46 & 2.59\\
\hline
HL estimator & 1.09 & 1.28 & 1.66 & 1.82 & 2.28 & 2.36\\ 
\hline
\end{tabular}
\end{adjustbox}
\end{center}

\vspace{0.1 in}

It is evident from the figures in Table 4 that ${\dot{\mbox{\boldmath $\mu$}}}_{\gamma,n}$ performs well in term of asymptotic efficiency for Cauchy distribution. For Spherical  distribution with $g(x)=e^{-x^{100}}$, it performs best among all the estimators. As expected, the sample mean performs best for normal distribution since it is the maximum likelihood estimator of location parameter in the normal distribution. The Hodges-Lehmann estimator is nearly as efficient as the sample mean for the normal distribution.\ On the other hand, the sample mean was outperformed by ${\dot{\mbox{\boldmath $\mu$}}}_{\gamma,n}$, co-ordinate wise median and Hodges-Lehmann estimator for Cauchy distribution since the sample mean does not have finite second moment when data follow Cauchy distribution.  

\vspace{0.2 in}

\noindent {\bf Acknowledgment:} The authors are thankful to the Editor-in-Chief\ Prof. Dipak Kumar Dey, an anonymous Associate Editor and an anonymous reviewer for many suggestions, which improved the article significantly. The authors would like to thank Prof.\ Marc Hallin for notifying them his paper and pointed out the condition assumed in Theorem 3.1.

\end{document}